\documentclass[preprint,sort&compress,12pt]{elsarticle}

\usepackage{graphicx}
\usepackage{amsmath}
\usepackage{bm}
\usepackage{lineno}
\usepackage{subfigure}
\usepackage{url}
\usepackage[colorlinks,unicode=true]{hyperref}

\journal{Nuclear Inst. and Methods in Physics Research, A}

\makeatletter
\def\ps@pprintTitle{
\def\@oddhead{\copyright\,2018. This manuscript is available under license CC BY-NC-ND 4.0 \hfill}
\def\@oddfoot{\it \small Preprint Accepted by Nuclear Inst. and Methods in Physics Research, A \hfill}}
\makeatother

\begin{document}

\begin{frontmatter}
\title{In-Orbit Instrument Performance Study and Calibration for POLAR Polarization Measurements}

\author[a1,a2]{Zhengheng~Li\corref{cor1}}
\ead{Zhengheng.Li@ihep.ac.cn}
\author[a3]{Merlin~Kole}
\author[a1]{Jianchao~Sun}
\author[a1]{Liming~Song}
\author[a4]{Nicolas~Produit}
\author[a1]{Bobing~Wu}

\author[a1]{Tianwei~Bao}
\author[a4]{Tancredi~Bernasconi}
\author[a3]{Franck~Cadoux}
\author[a1]{Yongwei~Dong}
\author[a1,a2]{Minzi~Feng}
\author[a4]{Neal~Gauvin}
\author[a5]{Wojtek~Hajdas}
\author[a1,a2]{Hancheng~Li}
\author[a1]{Lu~Li}
\author[a1,a2]{Xin~Liu}
\author[a5]{Radoslaw~Marcinkowski}
\author[a3]{Martin~Pohl}
\author[a6]{Dominik~K.~Rybka}
\author[a1]{Haoli~Shi}
\author[a6]{Jacek~Szabelski}
\author[a6]{Teresa~Tymieniecka}
\author[a1]{Ruijie~Wang}
\author[a1,a2]{Yuanhao~Wang}
\author[a1,a2]{Xing~Wen}
\author[a3]{Xin~Wu}
\author[a1]{Shaolin~Xiong}
\author[a6]{Anna~Zwolinska}
\author[a1]{Li~Zhang}
\author[a1]{Laiyu~Zhang}
\author[a1]{Shuangnan~Zhang}
\author[a1]{Yongjie~Zhang}
\author[a1,a7]{Yi~Zhao}

\address[a1]{Key Laboratory of Particle Astrophysics, Institute of High Energy Physics, Chinese Academy of Sciences, Beijing 100049, China}
\address[a2]{University of Chinese Academy of Sciences, Beijing 100049, China}
\address[a3]{University of Geneva (DPNC), quai Ernest-Ansermet 24, 1205 Geneva, Switzerland}
\address[a4]{University of Geneva, Geneva Observatory, ISDC, 16, Chemin d'Ecogia, 1290 Versoix, Switzerland}
\address[a5]{Paul Scherrer Institut, 5232 Villigen PSI, Switzerland}
\address[a6]{National Centre for Nuclear Research, ul. A. Soltana 7, 05-400 Otwock, Swierk, Poland}
\address[a7]{School of Nuclear Science and Technology, Lanzhou University, Lanzhou 730000, China}

\cortext[cor1]{Corresponding author.}

\begin{abstract}
POLAR is a compact space-borne detector designed to perform reliable measurements of the polarization for transient sources like Gamma-Ray Bursts in the energy range 50--500\,keV. The instrument works based on the Compton Scattering principle with the plastic scintillators as the main detection material along with the multi-anode photomultiplier tube. POLAR has been launched successfully onboard the Chinese space laboratory TG-2 on 15th September, 2016. In order to reliably reconstruct the polarization information a highly detailed understanding of the instrument is required for both data analysis and Monte Carlo studies. For this purpose a full study of the in-orbit performance was performed in order to obtain the instrument calibration parameters such as noise, pedestal, gain nonlinearity of the electronics, threshold, crosstalk and gain, as well as the effect of temperature on the above parameters. Furthermore the relationship between gain and high voltage of the multi-anode photomultiplier tube has been studied and the errors on all measurement values are presented. Finally the typical systematic error on polarization measurements of Gamma-Ray Bursts due to the measurement error of the calibration parameters are estimated using Monte Carlo simulations.
\end{abstract}

\end{frontmatter}



\section{Introduction}\label{sec:introduction}

Gamma Ray Bursts (GRBs) are short flashes of gamma-rays that appear in the sky at unpredicted times and from unpredictable directions. Since their discovery in the 1960s their exact nature is still not fully understood despite the properties of the prompt emission, like energy, time and direction having been measured in great detail by many other instruments~\cite{ZHANG2011206}. The polarization of the prompt emission is another important dimension that can help us understand the emission mechanisms of GRBs and the possible magnetic and geometric structure of the source~\cite{KUMAR20151, Beloborodov2017, Lyutikov2003}. Measuring the polarization  will allow to constrain different theoretical emission models. There have already been several attempts to measure the polarization of the high energy emission from GRBs by instruments such as the BATSE instrument onboard CGRO~\cite{Willis2005}, RHESSI~\cite{Coburn2003,Rutledge2004} and the IBIS and SPI instruments onboard INTEGRAL~\cite{Chauvin2013,Laurent2009}. All these instruments are however primarily designed for spectroscopy, timing and imaging, and therefore unoptimized for polarization studies. As a result their polarization measurements have large systematic and statistical errors. The small scale GAP detector onboard the IKAROS solar sail was the first dedicated GRB polarimeter which reported polarization measurements of three bright GRBs~\cite{Yonetoku2011, Yonetoku2012}. However in order to constrain emission models a larger sample of GRB polarization measurements with higher precision is required.

POLAR~\cite{PRODUIT2005616,Produit2017} is a compact space-borne detector with a wide field of view which is specially designed and optimized to measure the polarization of hard X-rays for transient sources like GRBs in the 50--500\,keV energy range. The POLAR experiment was proposed with the scientific goal to give a reliable polarization measurement of a large sample of GRBs~\cite{Xiong2009}. 

When the polarized photons interact with the detection materials through the Compton scattering process, the differential cross-section follows the Klein--Nishina equation~\eqref{equ:KN}.
\begin{equation}\label{equ:KN}
\begin{aligned}
\frac{d\sigma}{d\Omega} & =\frac{r_0^2}{2}\left(\frac{E'}{E}\right)^2\left(\frac{E'}{E}+\frac{E}{E'}-2\sin^2\theta\cos^2\phi\right) \\
& = \frac{r_0^2}{2}\left(\frac{E'}{E}\right)^2\left(\frac{E'}{E}+\frac{E}{E'}-\sin^2\theta+\sin^2\theta\cos\left(2\left(\phi+\frac{\pi}{2}\right)\right)\right)
\end{aligned}
\end{equation}
where $r_0$ is the classical electron radius, $E$ and $E'$ are the energies of the photon before and after the scattering process respectively, $\theta$ is the polar scattering angle and $\phi$ is the azimuthal scattering angle. After integration of $\theta$, the distribution of $\phi$ follows a distribution described by Eq. \eqref{equ:mc}
\begin{equation}\label{equ:mc}
f(\phi,E) = A(E)+B(E)\cdot\cos\left(2\left((\phi-\phi_0)+\frac{\pi}{2}\right)\right)
\end{equation}
where $\phi_0$ is the azimuthal scattering angle correlated to the polarization direction. The azimuthal scattering angle distribution of the scattered photons described by this function is called the modulation curve, and the ratio of $B$ over $A$ is called the modulation factor as presented by Eq. \eqref{equ:mcf} 
\begin{equation}\label{equ:mcf}
\mu(E) = \frac{B(E)}{A(E)}=\frac{f_{max}-f_{min}}{f_{max}+f_{min}}
\end{equation}
When the incident photons are $100\%$ linearly polarized $\mu$, for at a specific initial energy $E$, is defined as $\mu_{100}$. For partially polarized photons, the corresponding $\mu$ is in the range $0 - \mu_{100}$ and has a linear relationship with the polarization degree, then the polarization degree can be determined by $\Pi=\mu/\mu_{100}$. Based on this theory, the information of polarization including polarization degree and polarization direction can be directly measured by the distribution of the azimuthal scattering angle $\phi$ of the photons interacting through the Compton scattering process.

\begin{figure}[!ht]
\centering
\subfigure[]{\includegraphics[height=5cm]{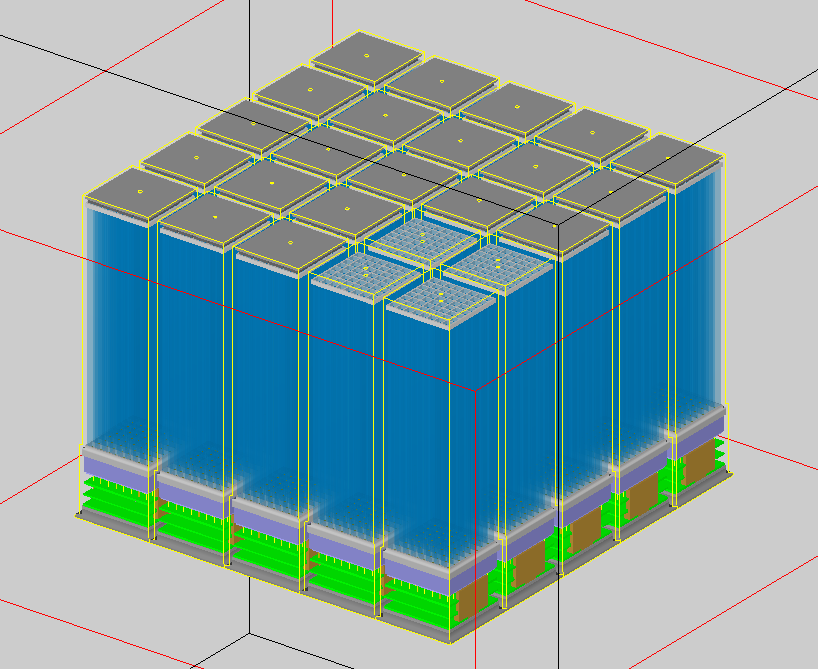}}
\hspace{2mm}
\subfigure[]{\includegraphics[height=5cm]{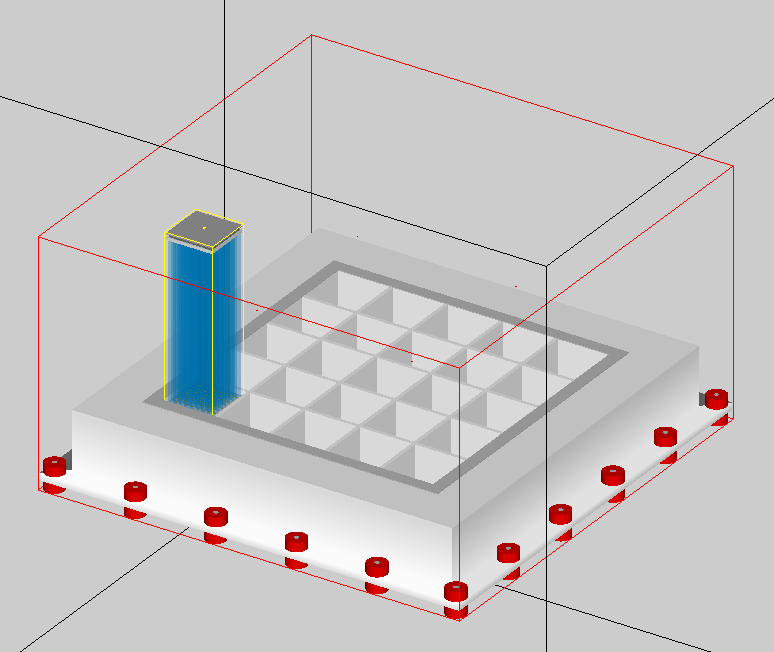}}
\caption{Geometric structure of 25 modules with the end plate and dampers of 4 modules removed for clarification (a) and the aluminum frame of POLAR OBOX with one module installed (b) shown separately in the GEANT 4 mass model without CFRP.}
\label{fig:obox_geo_struct}
\end{figure}

POLAR measures the Compton scattering angles using plastic scintillator (PS) bars with the type EJ-248M \cite{EJ248M}. This material was chosen for the reason of its low-z characteristic which contributes to a higher Compton scattering cross-section in the energy range of POLAR, as well as its higher temperature resilience. As shown in Figure~\ref{fig:obox_geo_struct}, 64 PS bars of dimension $5.8\times 5.8 \times 176\,mm$ are grouped together as an $8 \times 8$ array and a 64-anodes photomultiplier tube (MAPMT) from Hamamatsu with its corresponding front-end electronics (FEE) connected to this PS array is used to readout and process the signals. Each PS bar is separated by a piece of thin highly reflective film of the type Vikuiti Enhanced Specular Reflector Film (ESR) \cite{3M} to increase the fluorescence photons collection and reduce the optical crosstalk. This structure with some other components such as the outer Carbon Fiber Reinforced Polymer (CFRP) shell forms a standalone module and 25 such modules are installed in an aluminum frame as a $5 \times 5$ array. Such a PS detection plane with $40 \times 40$\,pixels enables POLAR to precisely measure the projection of the azimuthal scattering angle distribution. The 25 modules are connected by a central trigger (CT) system installed inside the aluminum frame which is responsible for implementing the overall trigger logic, collecting and packaging event data, etc. A more detailed description of the design and construction of the POLAR flight model is provided in Ref.~\cite{Produit2017}. 

POLAR was launched onboard the Chinese space laboratory TG-2 on 15th September in 2016 and successfully switched on afterwards. More than 50 GRBs have been detected by POLAR and confirmed by other satellites such as Fermi GBM, Swift, etc. during the first 6 months of operation \cite{Xiong2017}. POLAR also detected strong signals from the Crab Pulsar \cite{Hancheng2017} and Solar Flares. Before the polarization analysis of these detected GRBs and other sources, a full study of the instrument performance during in-orbit operation is required and all the related in-orbit calibration parameters of the instrument itself should be provided. These in-orbit calibration parameters include pedestal and noise levels of the FEE, the nonlinearity function of the gain in the electronics, the ADC threshold of each channel, the crosstalk matrix of each module and the gain of each channel. All these calibration parameters are not only necessary in the analysis pipeline to reconstruct the deposited energy of each bar and reduce the systematic effects from the instrument, but also work as the input parameters for the Monte Carlo simulations~\cite{Kole2017} that simulate the digitization process of the PMTs and FEEs and handle the event response, which is another important procedure of the polarization analysis by comparing the modulation curves between the measured data and the simulated data. Therefore, the methods to calculate all these calibration parameters using in-orbit data and their typical values are firstly discussed and presented. It is found in the in-orbit data of POLAR that there is a certain percentage of fake events that are generated by FEEs possibly induced by the background high energy charged particles. Because the contamination of those fake events will affect the calibration, a method to filter those fake events is also provided.

Unlike the on-ground experiment, the space environment, including temperature, different density of charged particles etc. is complicated and always varies over time while POLAR arrives at different positions in-orbit. During the first 6 months of operation, the instrument settings were also changed several times for different purposes. The most important change is the high voltage (HV) setting for the purpose of gain--HV relation measurement. Therefore, both the effect of temperature change on all calibration parameters were carefully checked and studied as well as the relationship between gain and high voltage. The relations between gain and high voltage were studied using in-orbit data for the purpose of calculating the gain parameter corresponding to the high voltage setting for the normal scientific data acquisition. Finally, as all the calibration parameters have measurement errors, and they will result in a systematic error on the final modulation curves, the typical level of the calibration parameter induced systematic error on the polarization measurement of GRBs was studied and evaluated by Monte Carlo simulation.

\newpage

\section{Methods to measure the calibration parameters}\label{sec:calib_pars}

The analysis chain for the in-orbit data of POLAR can be described in general by: subtraction of pedestal and common noise $\Rightarrow$ data filtering $\Rightarrow$ gain nonlinearity correction $\Rightarrow$ threshold calculation $\Rightarrow$ crosstalk correction $\Rightarrow$ energy calibration. The details and the methods to measure all the calibration parameters that are needed by each step of this analysis chain will be discussed in this section. Firstly, a short description of the DAQ process and the trigger logic of POLAR will be given which is needed for better understanding the discussion of the methods to measure all the calibration parameters. Then the methods to measure different calibration parameters will be discussed subsequently in different subsections.

\subsection{DAQ process and trigger logic}\label{sec:daq_process}

\begin{figure}[!ht]
\centering
\includegraphics[width=8cm]{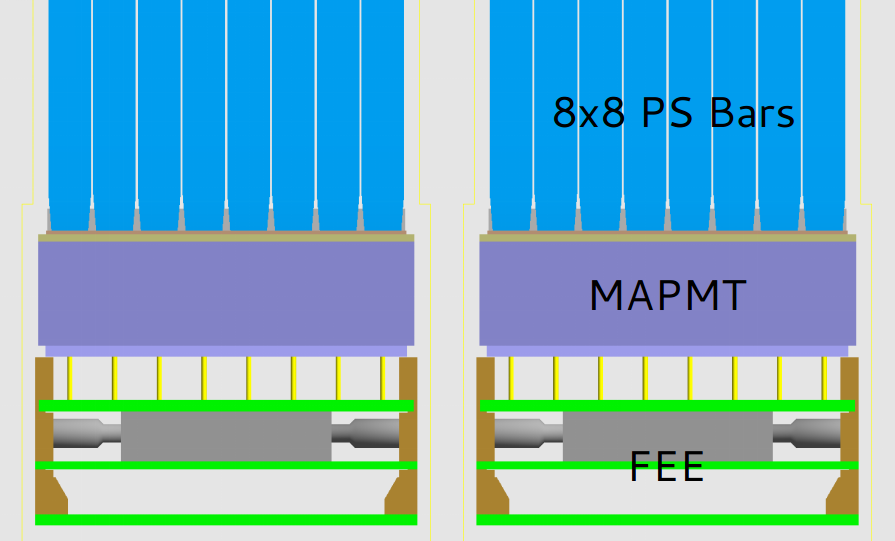}
\caption{A more detailed geometric structure of the bottom side of two adjacent modules.}\label{fig:modules}
\end{figure}

Figure~\ref{fig:modules} shows a more detailed geometric structure of the bottom part of two adjacent modules. For hard X-rays there is a high possibility for Compton scattering to occur in one PS bar followed by one or several interactions in other different bars, these bars can be in different modules. The deposited energy of each bar will be firstly converted to fluorescent optical photons then collected at the bottom of the bar. Even though the shape of the end of each bar is made pyramid-like and there is the baffle placed between two adjacent bars as shown in Figure~\ref{fig:modules}, the photon crosstalk between different bars can still happen in the area of the optical coupling film and the PMT window, whose total thickness is about $1.5\,mm$. It is therefore possible for bars without an energy deposition to have a signal in the corresponding channels. This signal can also be larger than the trigger threshold inducing a false trigger signal. Photoelectrons will be produced on the photocathode by the collected optical photons from real deposited energy, or crosstalk, then be amplified by the 12 dynodes in the MAPMT in each channel. The FEE at the bottom of each module handles the readout and processes the signal of each channel from the corresponding MAPMT. The process mainly includes taking trigger decision based on the threshold setting, communicating with the CT, converting the analog signal to an ADC value, tagging a time stamp and packaging the digital data into a certain format.

A physical event can occur within one module and can also occur across several different modules within a coincidence time window of about $3\,\mu s$. Only multi-scattering events, that is events with 2 or more triggering channels, are useful for polarization measurements. A specific trigger logic is implemented in the CT for the purpose of saving data space, suppressing background and reducing the dead time of the instrument. Once a module is triggered by at least one channel, four additional flags will be issued by the FEE to indicate:

\begin{itemize}
\item if there is at least one channel triggered (TOUT1) in this module
\item if there are at least two channels triggered (TOUT2) in this module
\item if there are too many channels triggered (TOO-MANY) in this module
\item if the intensity of the sum of the signal from all channels in the module is too high (TOO-HIGH).
\end{itemize}

High energy charged particles like protons and electrons as well as cosmic ray ions are capable of penetrating the full detector where the ionization process will result in a large number of triggering channels. The total deposited energy, measured in ADC, will also be high. Therefore the TOO-MANY and TOO-HIGH flag can tag this kind of event to a certain degree. The CT implements its trigger logic based on these four flags of the 25 modules. Based on the information from the modules the CT can take the following decisions:

\begin{itemize}
\item In case only one module is triggered and only the TOUT1 flag of the module is true, the event will be tagged as a `single event'
\item If one event has at least one module triggered and with true TOO-MANY or TOO-HIGH flag, this event will be tagged as a `cosmic event'. 
\item All other cases of triggered events are tagged as a `normal event' in the CT. 
\end{itemize}

All normal events are stored whereas only part of the single events and cosmic events are kept using a prescale logic in order to study the quality of the online event selection. The prescale values which determine the ratio of the single and cosmic events to be stored can be programmed in the CT and can be altered during flight. A more detailed description about the trigger logic of POLAR is provided in Ref.~\cite{Rybka}.

\subsection{Gain nonlinearity of electronics}\label{sec:non_linear}

\begin{figure}[!ht]
\centering
\subfigure[]{\includegraphics[height=5.1cm]{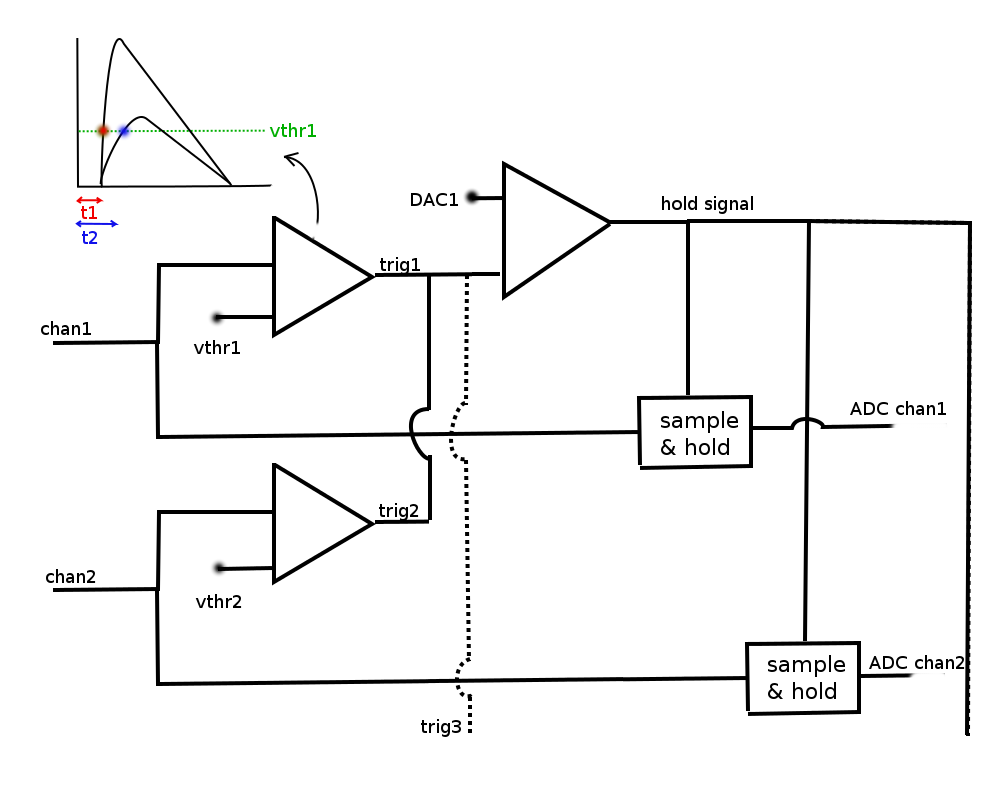}}
\hspace{2mm}
\subfigure[]{\includegraphics[height=5.1cm]{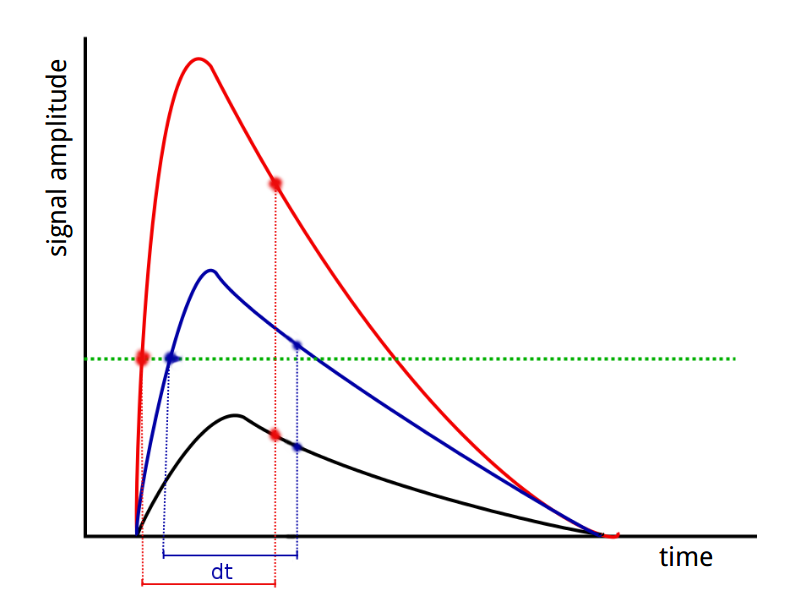}}
\caption{Explanation of the gain nonlinearity of electronics. (a) and (b) are respectively the schematic representation of the trigger logic and the signal sampling and holding inside the FEE.}\label{fig:trigger_logic}
\end{figure}

The 64 signals of a physical event from the PMT channels are each compared with an individual threshold level in the threshold comparator within a module. As shown in Figure~\ref{fig:trigger_logic}(a), the threshold comparators of any channels whose signal exceeds the threshold level will send a signal to a second comparator, the multiplicity comparator, which is in charge of counting the number of triggered channels. The multiplicity comparator will issue a signal, referred to as `DAC1', at the time when the first channel has triggered. The signal waveforms of all 64 channels within the module will be sampled and held at a same fixed delay time $dt$ after the DAC1 is issued. Then the CT will decide whether to digitize and store the 64 held signals of this event according to the trigger logic described in Section~\ref{sec:daq_process}. 

\begin{figure}[!ht]
\centering
\subfigure[]{\includegraphics[height=4.7cm]{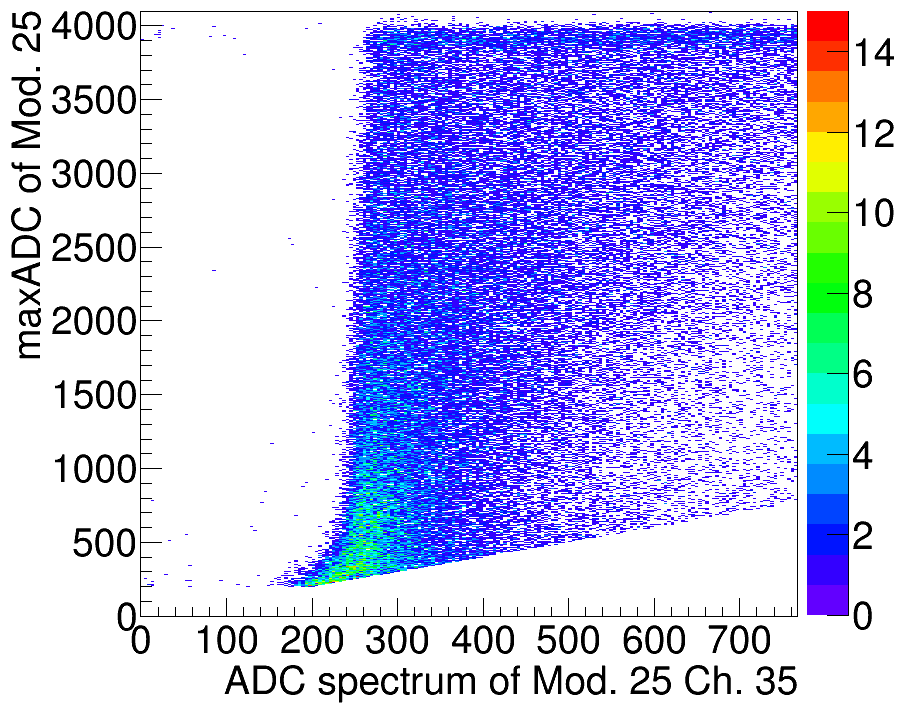}}
\hspace{3mm}
\subfigure[]{\includegraphics[height=4.7cm]{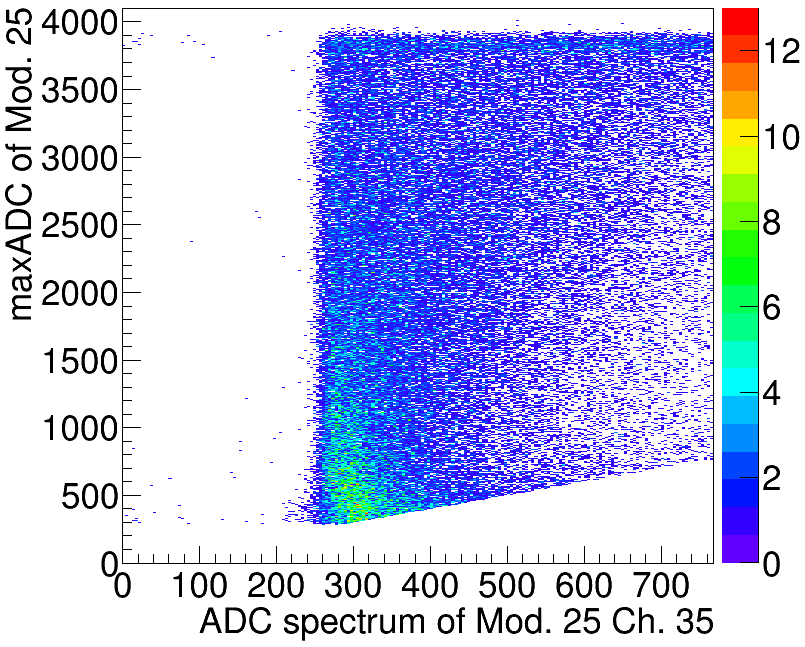}}
\caption{Relation between the measured ADC spectrum of module 25 channel 35 and the maximum ADC value measured in module 25 before (a) and after (b) the gain nonlinearity correction, where the value of Z-axis is the event counts.}\label{fig:maxADC}
\end{figure}

As shown in Figure~\ref{fig:trigger_logic}(b), for the FEE of POLAR the time at which the waveforms are held is slightly after the peak in the waveforms. Therefore, the sampling point will be closer to the peak when the sampling time is earlier, as a result the readout signal will be larger. As the delay time $dt$ is fixed, the sampling time of all 64 channels within one module is actually determined by the time when the DAC1 is issued, and the DAC1 issuing time is determined by the time when the signal passes the trigger threshold. It is known that the rise time of a signal with a larger amplitude tends to be faster, which means that the triggering time of the channels with larger signal will trigger earlier. Thus, the DAC1 issuing time will be determined by the largest signal of the 64 channels in one module, which means the readout ADC of the 64 channels will have a dependence on the maximum ADC value in the module for one event. An example of this dependence is shown in Figure~\ref{fig:maxADC}(a). This figure shows the relation between the ADC spectrum of one channel from the triggering events and the maximum ADC value, referred to as maxADC, measured in the module where the channel is in. In the figure the events where the maxADC is of the channel itself are not included. For low values of the maxADC it can be seen that the spectrum of the channel is shifted significantly towards lower ADC values as a result of the relatively later signal sampling time of the waveform.

\begin{figure}[!ht]
\centering
\includegraphics[width=8cm]{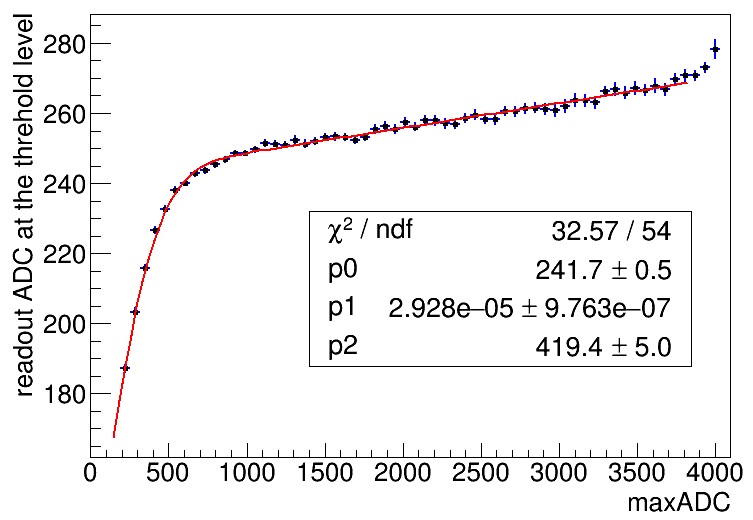}
\caption{The readout ADC value at the threshold level for module 25 channel 35 as a function of the maximum ADC value in module 25.}\label{fig:non-linear}
\end{figure}

A curve showing the nonlinear behavior of the electronics, which depends on the maxADC, can be measured for each channel by measuring the ADC value at the threshold level, that is the position of the left cutoff of the ADC spectrum of the channel as shown in Figure~\ref{fig:maxADC}(a), as a function of the maxADC value in the module. An example of the nonlinearity curve is shown in Figure~\ref{fig:non-linear}. The nonlinearity factor as a function of the maxADC can be well fitted by Eq.~\eqref{equ:non-linear}.
\begin{equation}\label{equ:non-linear}
f(maxADC) = \frac{1}{2}p_0\left(1 + p_1 \times maxADC\right) \left(1 + \mathrm{erf}\left(\frac{maxADC}{p_2}\right)\right)
\end{equation}
This function can be normalized at a specific point of the maxADC and subsequently used to correct the gain nonlinearity of the electronics for the readout ADC of each channel by Eq.~\eqref{equ:non-linear-corr}.
\begin{equation}\label{equ:non-linear-corr}
ADC_{lin} = \frac{ADC_{nonlin}}{f(maxADC) / f(maxADC_0)}
\end{equation}
where $ADC_{nonlin}$ is the measured nonlinear ADC value of one channel, and $ADC_{lin}$ is the linear one after the correction by the nonlinearity function normalized at $maxADC_0$. Figure~\ref{fig:maxADC}(b) shows the relation between the ADC spectrum of one channel and the maximum ADC value in the module after the correction for the gain nonlinearity. It can be seen that the ADC value at the threshold level for that channel is almost flat after the gain nonlinearity correction.

\subsection{Pedestal and Noise}\label{sec:pedestal_noise}

Apart from the three types of event introduced in Section~\ref{sec:daq_process}, there is the fourth type of event in data, which is `pedestal event'. Pedestal events are non-physical events but are instead the result of a forced trigger of all channels on all 25 modules by the CT. A pedestal event is taken every second. Every channel has a none-zero baseline signal even when there is no signal input from the PMT. The baseline signal differs among different channels and may also change under different test conditions like temperature. The 1Hz pedestal event readout rate has the purpose of measuring and tracking the baseline signal of each channel with respect to the operating temperature along the POLAR orbit. Pedestal events are also used to track the difference of the time stamp counter between the CT and the 25 FEEs which is important for performing event time alignment between the two different kinds of data packets from the CT and the 25 FEEs. A small possibility exists for physical events to occur and deposit energy in one bar during the coincidence time window when the FEE is processing a pedestal event, the probability of this is however negligible ($\ll1\%$) for the typical event rates encountered in orbit.

\begin{figure}[!ht]
\centering
\subfigure[]{\includegraphics[height=3.7cm]{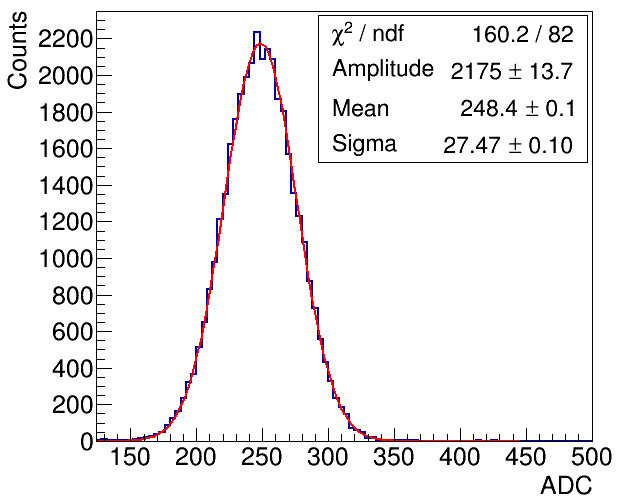}}
\subfigure[]{\includegraphics[height=3.7cm]{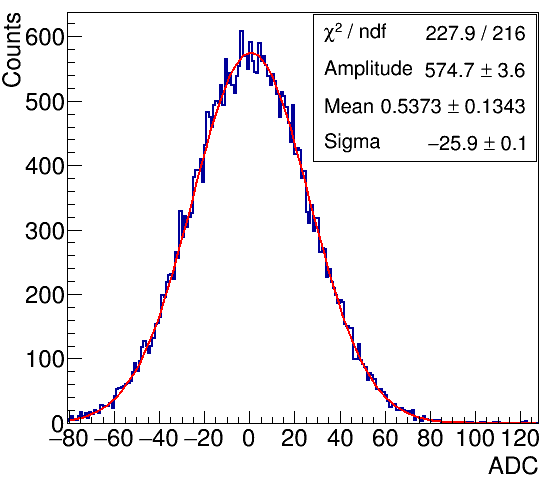}}
\subfigure[]{\includegraphics[height=3.7cm]{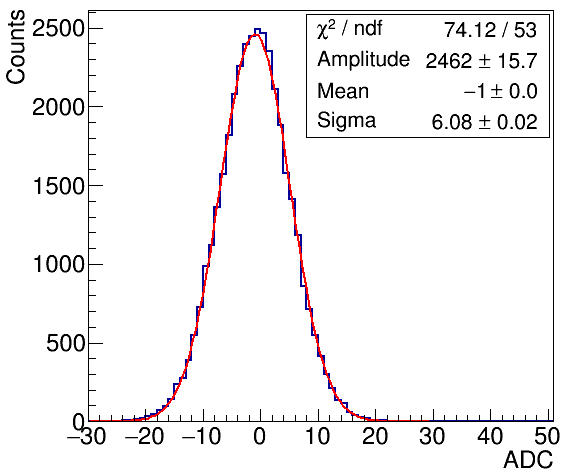}}
\caption{Pedestal and noise of module 05 channel 00 at $25^\circ C$. (a) is the distribution of the pedestal ADC value of the channel, (b) is the distribution of the common noise of the module and (c) is the distribution of the intrinsic noise of the channel.}\label{fig:pedestal}
\end{figure}

After accumulating a large number of pedestal events for each channel, the distribution of the measured pedestal ADC values can be well fitted by a Gaussian function. The value of the mean and the width can be acquired as shown in Figure~\ref{fig:pedestal}(a). The mean value of the distribution needs to be subtracted from the ADC values measured for physical events. The variation of pedestal ADC values ($\sigma_t$) is caused by the electronic noise either from the FEE itself or injected into the FEE. The noise in POLAR can be categorized into two types, the intrinsic noise ($\sigma_i$) and the common noise ($\sigma_c$). The intrinsic noise is the component of the readout noise of each channel which is uncorrelated to that measured in the other channels in the module. The common noise is the correlated signal shift of all the 64 channels of one module with the same direction and the similar amplitude.

\begin{figure}[!ht]
\centering
\includegraphics[width=7.5cm]{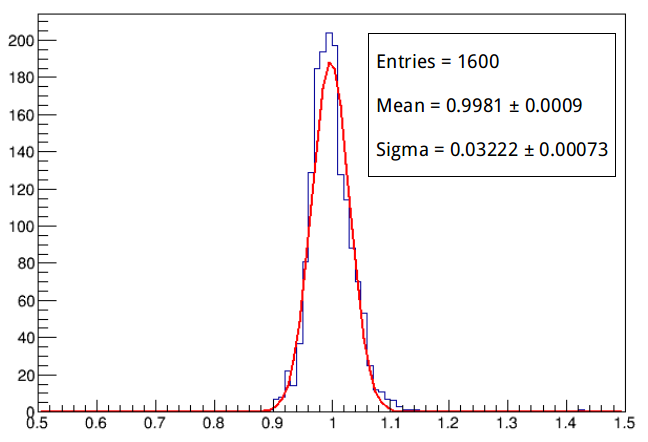}
\caption{Distribution of $\sqrt{\sigma_i^2+\sigma_c^2}/\sigma_t$ of all 1600 channels.}\label{fig:error_ratio}
\end{figure}

The common noise can be estimated and subtracted for each event in one module by calculating the mean ADC value of all the channels without trigger and not adjacent to the triggering channels for reducing the contamination coming from the crosstalk signal of energy deposition. Figure~\ref{fig:pedestal}(b) shows the distribution of the common noise of one module which is fitted by a Gaussian function. The level of the intrinsic noise of one channel can be acquired by refitting the distribution of ADC values of the channel using pedestal events after subtracting the pedestal mean value and the common noise as shown in Figure~\ref{fig:pedestal}(c). The width of the pedestal ADC value, that is the total noise, should be approximately the sum of intrinsic noise and common noise, so $\sigma_t^2 \approx \sigma_i^2 + \sigma_c^2$. Figure~\ref{fig:error_ratio} shows the distribution of the ratio of $\sqrt{\sigma_i^2+\sigma_c^2}$ over $\sigma_t$ of all 1600 channels, from which it can be seen that the mean ratio is almost 1 as expected. The typical value of the intrinsic noise is less than 10\,ADC, while for common noise it is about 30\,ADC, the total ADC range being from 0 to 4095 ADC channels. As the common noise is calculated as the mean of the ADC shift of the non-triggering channels in one module, the typical measurement error of the common noise is $\sim \langle\sigma_i\rangle/\sqrt{64}$. As a result the remaining noise on each channel after the subtraction of common noise is therefore of the order of $\sigma_i+\langle\sigma_i\rangle/\sqrt{64}$. 

\subsection{Data filtering}\label{sec:data_filtering}

\begin{figure}[!ht]
\centering
\subfigure[]{\includegraphics[height=4cm]{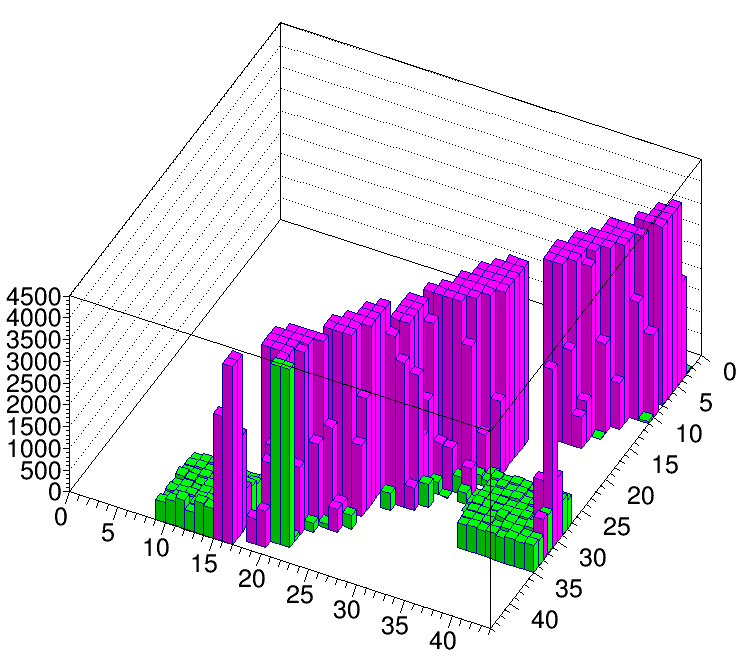}}
\subfigure[]{\includegraphics[height=4cm]{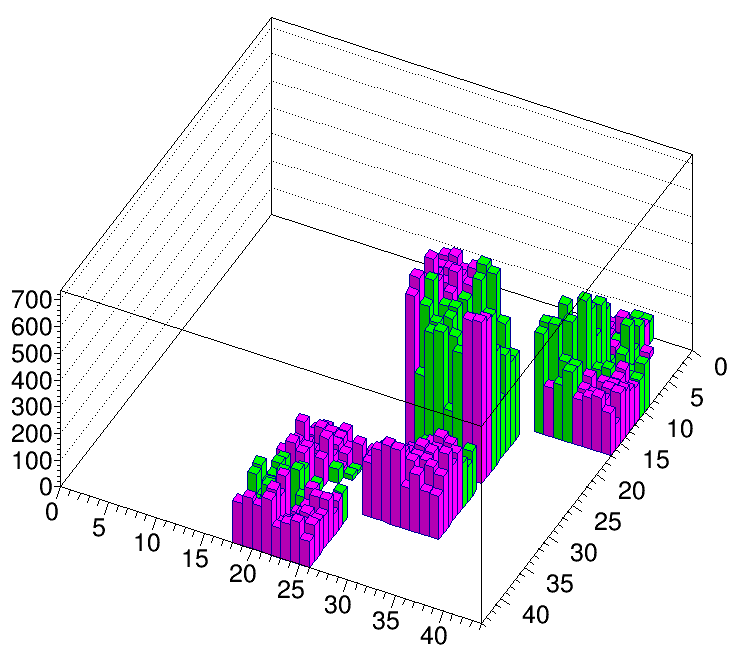}}
\subfigure[]{\includegraphics[height=4cm]{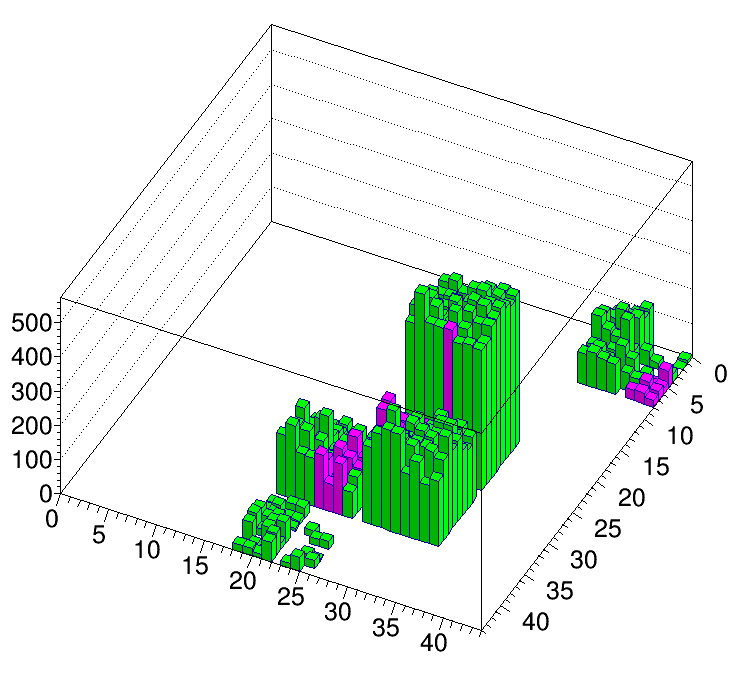}}
\caption{The 3D view of three events adjacent in time (a--c) induced by a single cosmic ray. The X and Y-axis are the column and row number of the channels in the instrument, while the Z-axis shows the raw ADC value recorded by each channel. Triggering channels are shown as purple (or dark gray) while non-triggering channels are green (or light gray).}\label{fig:post_cosmic}
\end{figure}

The in-orbit data of POLAR shows a non-negligible amount of abnormal events typically characterized by all the triggering channels having ADC values below their threshold level. The ratio of these abnormal events out of the total number of events was found to increase with magnetic latitude. Furthermore, it was found that the great majority of the events triggering directly after cosmic events show such characteristics, indicating that the abnormal events may be induced by the effects of cosmic rays on the FEE. Figure~\ref{fig:post_cosmic} illustrates this effect by showing the hit patterns of three consecutive events, where the purple bars indicate triggering channels while the green bars indicate non-triggering channels. The height of the bar indicates the raw ADC value recorded by each channel. The first event with a line-shape hit pattern and most of the triggering channels saturating at the ADC range limit (4095) is due to a charged particle with high energy traversing the full instrument. It is correctly identified by the trigger logic as being a cosmic event based on the large number of triggering channels and the high intensity of the signal. After this event, all the FEEs of the modules traversed by the charged particle show abnormal behavior for a short time and generate several non-physical events. The time between these events is of the order of the absolute dead time of the FEE ($\sim 65\,\mu s$), indicating that the events are likely connected to the first event and that these events result from the FEE's recovering from the large energy depositions. It can be seen that a large number of the channels in the modules which are not triggering in the first cosmic event are triggering in the second event but with very low ADC values. The second event is however not tagged as a cosmic event in the CT, therefore it is recorded. The third event appears also abnormal resulting from the first cosmic event with several of the channels in those modules triggering with low ADC values. The third event is also not tagged as a cosmic event and recorded. The event following the third event is normal again and is separated in time with a period significantly longer than the dead time.

The contamination of those non-physical events will affect several stages of the in-orbit calibration. Due to their clear characteristics, however, it is possible to filter out these abnormal events from physical events by applying relatively simple cuts. This filtering is applied in the analysis chain immediately after the calculation and subtraction of the pedestal mean value and the common noise. 

\begin{figure}[!ht]
\centering
\includegraphics[width=8cm]{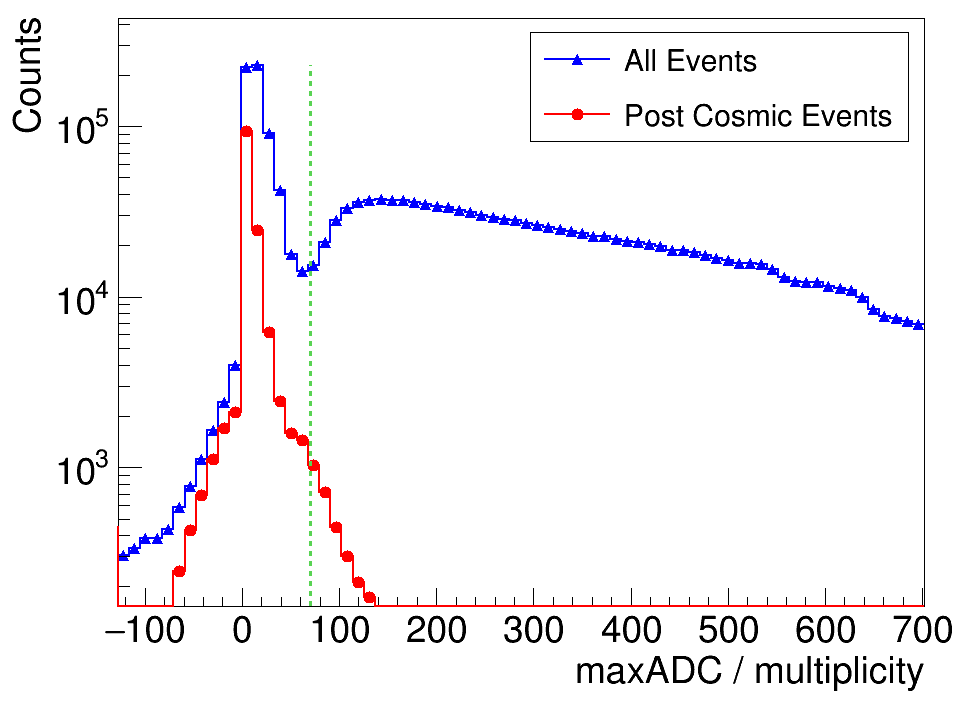}
\caption{Distribution of the maximum ADC of the triggered channels divided by the trigger multiplicity of a single module from in-orbit data. The distribution of all events is shown in blue (triangle), the distribution from events after cosmic events with waiting time $<100\,\mu s$ is shown in red (circle). The green dashed line shows the cut value used in data filtering, which is 70\,ADC. By this cut value more than $95\%$ of the post cosmic events can be selected out.}\label{fig:max_adc_dm}
\end{figure}

\begin{figure}[!ht]
\centering
\subfigure[]{\includegraphics[height=4.7cm]{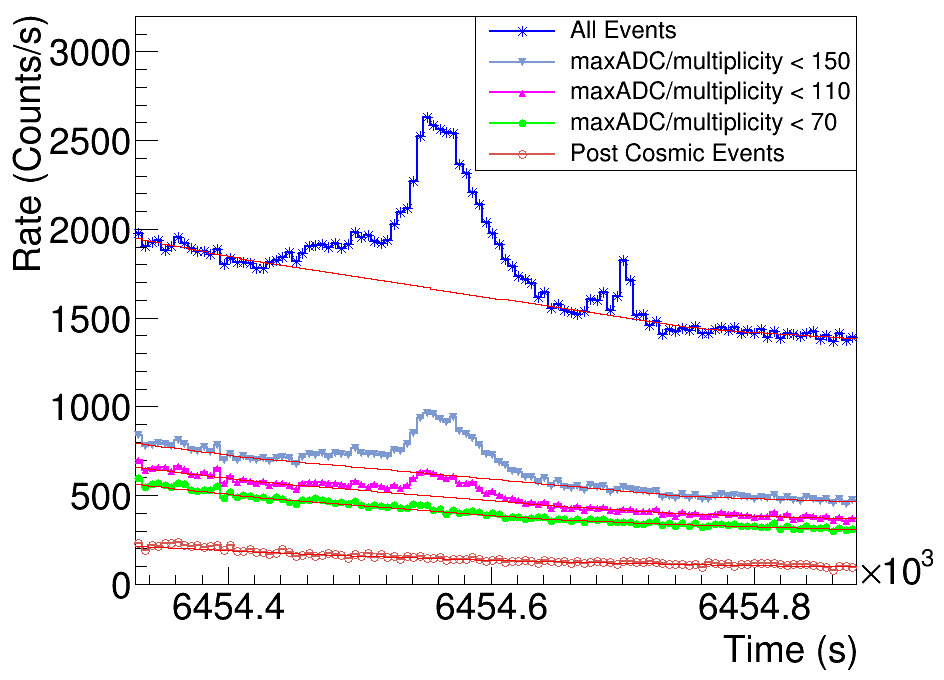}}
\hspace{1mm}
\subfigure[]{\includegraphics[height=4.7cm]{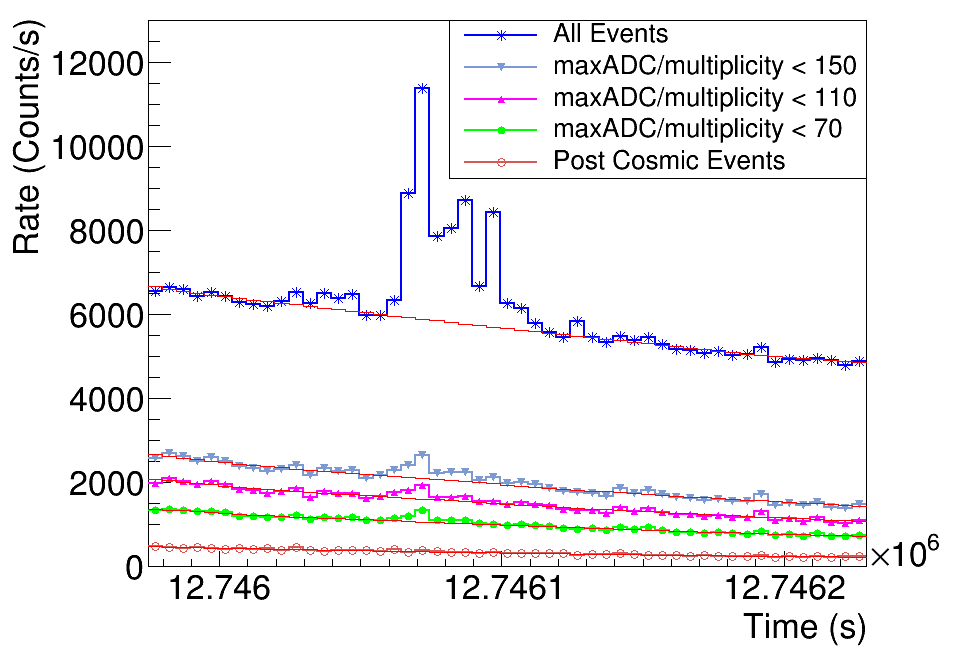}}
\caption{Rate curves of the abnormal events filtered by three different cut values of the maxADC/multiplicity and the rate curves of all events and the post cosmic events of a solar flare event and two GRB events detected by POLAR. (a) is of the solar flare event with a short time GRB event (GRB~161129A) just after it, (b) is of a long time duration GRB event (GRB~170210A) with the single bar detection mode.}\label{fig:data_filtering}
\end{figure}

The abnormal events can be found in the data by checking if they directly follow cosmic events (with true TOO-MANY or TOO-HIGH flag in the module level) and have short time between consecutive events ($<100\,\mu s$). This method, however, can only find a small fraction of those abnormal events because most of the cosmic events are discarded during flight due to the prescale logic while those abnormal events induced by those discarded cosmic events are not tagged as cosmic and therefore recorded. Thus, it is impossible to identify all the abnormal events only based on the recorded cosmic events and the time between events. Figure~\ref{fig:max_adc_dm} shows the distribution of the maximum ADC of the triggered channels, which is after pedestal and common noise subtraction, divided by the trigger multiplicity in one module for both all events and the events directly following cosmic events. The events following cosmic events (post cosmic events) are selected using the criteria that the preceding event is a cosmic event and that the time between the events is $<100\,\mu s$. There is a clear valley at around 70\,ADC found in the distribution of the maxADC/multiplicity value of all events for all modules, and most of the values of maxADC/multiplicity of the post cosmic events are smaller than 70\,ADC ($>95\%$). This distribution shows the value of maxADC/multiplicity of one module to be a good characteristic value to identify the abnormal events shown in Figure~\ref{fig:post_cosmic}(b--c). This cut based on maxADC/multiplicity can therefore be used to remove these post cosmic events even if the preceding cosmic events are not recorded. However the value of the cut on the maxADC/multiplicity should be chosen carefully. Figure~\ref{fig:data_filtering} shows the light curves of a solar flare event and two GRB events together with the rate curves of the post cosmic events and the rate curves of the filtered events by 3 different cut values of maxADC/multiplicity. The applied cut values are 70, 110, 150\,ADC respectively. As expected, the ratio of the valid photon events that are filtered out by mistake with the cut according to the maxADC/multiplicity is increasing with the cut value. From Figure~\ref{fig:max_adc_dm} it can be seen that with the cut value 70\,ADC more than $95\%$ of the abnormal post cosmic events can be removed. While for the valid photon events the removed amount by mistake using this cut value is less than $5\%$ which can be seen from the rate curves shown Figure~\ref{fig:data_filtering}. For performing the calibration of POLAR this cut value is acceptable, for future polarization measurements the cut value will be optimized for each GRB event in order to maximize the signal to background ratio.

\subsection{Threshold calculation}\label{sec:threshold_calculation}

The threshold level of each channel in POLAR is determined by two different threshold settings in the FEE. One is a general threshold that is shared among all channels in the same FEE of one module. The other one is an offset value for each channel that is relative to the general one, which can be slightly adjusted for different channels. Therefore the final threshold in the FEE for each channel is the general one of the module plus the offset of each channel in the module. The offset setting is used to tune the threshold level for each channel and reduce the threshold non-uniformity to get a flatter detection efficiency among all channels which is important for polarization measurements. 

\begin{figure}[!hb]
\centering
\subfigure[]{\includegraphics[height=4.6cm]{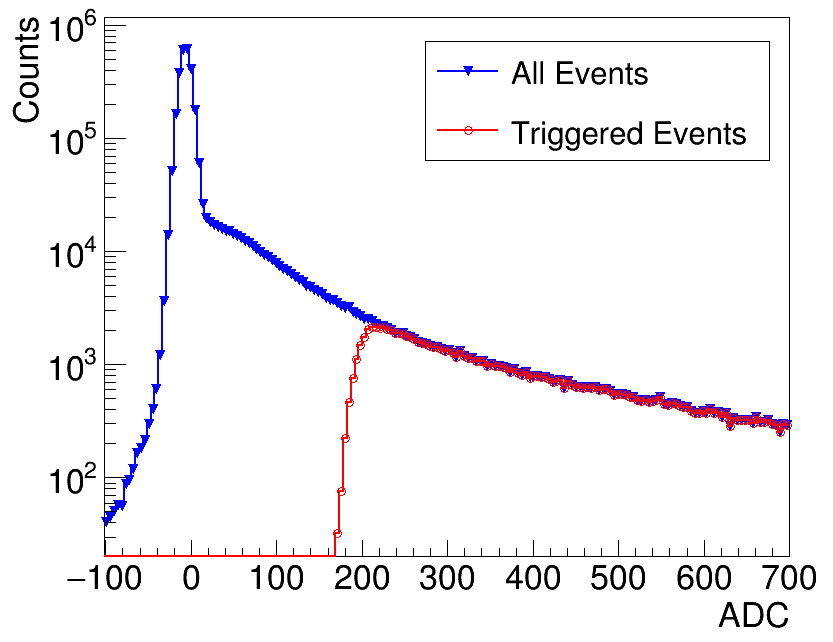}}
\hspace{2mm}
\subfigure[]{\includegraphics[height=4.6cm]{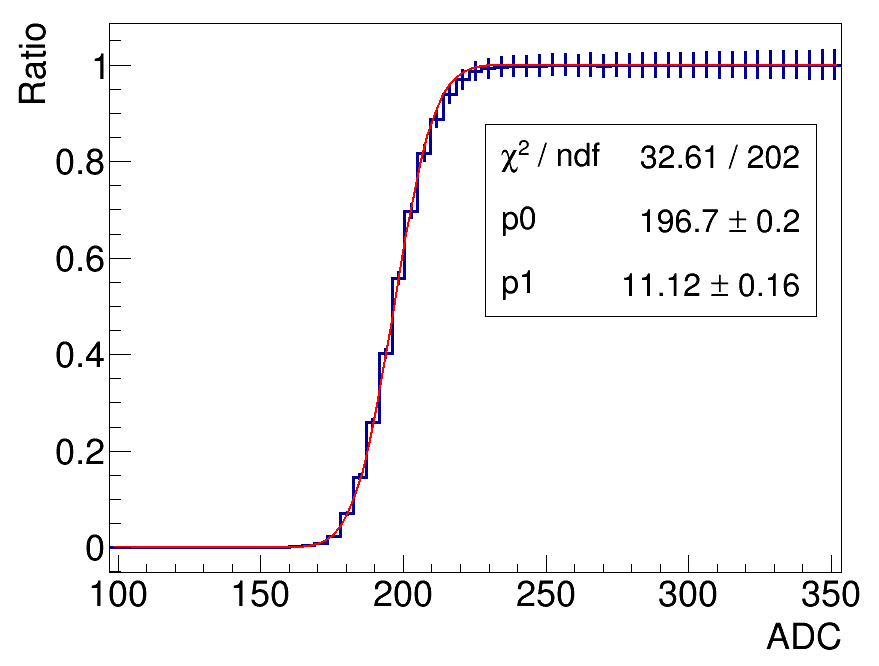}}
\caption{Threshold calculation for module 05 channel 10. (a) is the ADC spectra of the triggered events (red and circle) and all events (blue and triangle) for the channel, from the events where at least two in the other 63 channels are triggered, (b) is the the ratio curve of the spectrum of the triggered events divided by the spectrum of all events.}\label{fig:threshold}
\end{figure}

The threshold level in the FEE of each channel is firstly measured in unit of ADC. The energy threshold in unit of keV can then be calculated after the measurement of gain. The measurement of ADC threshold for all 1600 channels needs to be performed from data properly for the purpose of tuning the threshold setting in the FEE and for later use in the Monte Carlo simulation. As discussed in Section~\ref{sec:non_linear}, the readout ADC value of each channel in the FEE corresponding to the input signal from PMT has a nonlinear property which is dependent on the maxADC in one module, the gain nonlinearity should be corrected before the threshold calculation, otherwise the calculated ADC threshold level will also be dependent on the maxADC value in the module. The ADC threshold position of each channel is therefore calculated using the data after removing the cosmic ray induced non-physical events, after pedestal and common noise subtraction and also after the gain nonlinearity correction for each channel. The mean position and width of the ADC threshold of one channel can be measured by firstly producing an ADC spectrum of all events for the channel and a second spectrum only containing the events when the channel is triggered, then dividing the two spectra to get a curve of the ratio of the second one over the first one as shown in Figure~\ref{fig:threshold}. The ratio value of each bin in the curve is actually the percentage of the triggered events out of all events in the ADC range of the bin. It should be noted here that the effect of the data loss of single bar events caused by the prescale logic in the CT as discussed in Section~\ref{sec:daq_process}, and also of the not recorded events which have small energy deposition in the module but with no channel triggered, should be taken into account when calculating the ratio. The data loss of single bar events and the non-triggered events will reduce the counts in the spectrum of all events at low ADC regime around the threshold position, which will result in a bigger ratio than the real one. Consequently, this will result in the measured mean position of the ADC threshold to be underestimated and the width to be overestimated. This effect can be avoided when calculating the ADC threshold of one channel by only selecting events where at least two in the other 63 channels are triggered, independent of the trigger status of the current channel. The two ADC spectra shown in Figure~\ref{fig:threshold}(a) are therefore accumulated from those events satisfying this selection criteria based on the trigger multiplicity. The ratio curve shown in Figure~\ref{fig:threshold}(b) can be well fitted by the function presented in Eq.~\eqref{equ:threshold}.

\begin{equation}\label{equ:threshold}
f(E) = \frac{1}{2}\left(1 + \mathrm{erf}\left(\frac{E - p_0}{\sqrt{2}p_1}\right)\right)
\end{equation}
where $E$ is the ADC value of the channel, $p_0$ is the measured mean ADC threshold position, the typical value of which for the normal data acquisition setting when in orbit is about 200\,ADC, $p_1$ is the width.

\begin{figure}[!ht]
\centering
\subfigure[]{\includegraphics[height=4.5cm]{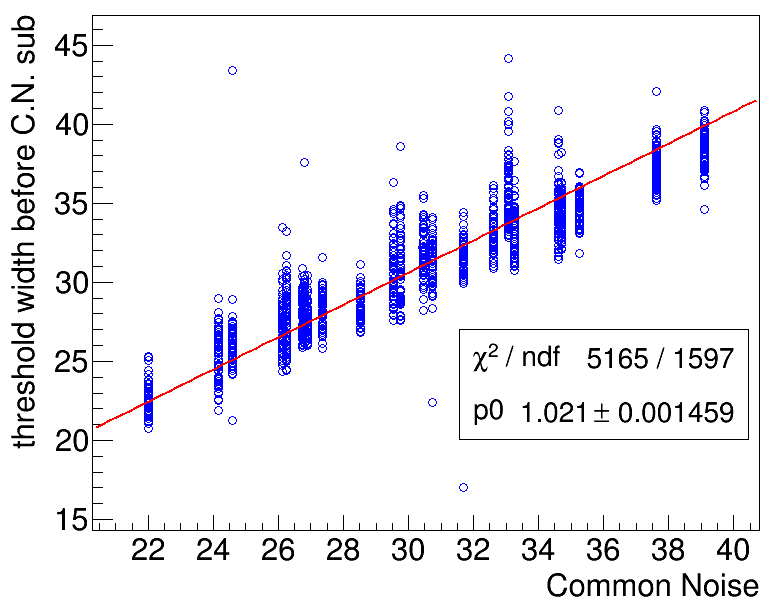}}
\hspace{2mm}
\subfigure[]{\includegraphics[height=4.5cm]{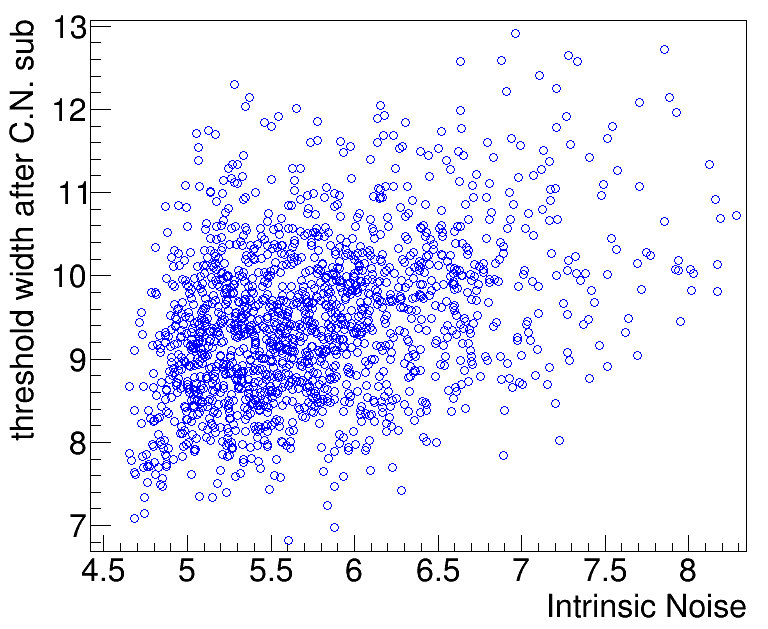}}
\caption{The correlation between the threshold width and noise. (a) shows the threshold width calculated from data before common noise subtraction vs. common noise and (b) shows the threshold width calculated from data after common noise subtraction vs. intrinsic noise.}\label{fig:threshold_width_noise}
\end{figure}

The ADC threshold position and width can also be calculated using the data before common noise subtraction, that is, using the gain nonlinearity corrected data with the common noise added back for each event. It is found that the width of the threshold is larger when it is calculated using the data before common noise subtraction. Furthermore a clear linear correlation is found between the threshold width and the common noise width of the module, as shown in Figure~\ref{fig:threshold_width_noise}(a). Whereas, no correlation is found between the threshold width calculated using data after common noise subtraction and the intrinsic noise width of each channel, as shown in Figure~\ref{fig:threshold_width_noise}(b). This indicates that the common noise should be injected into the signal after the trigger decision, thereby enlarging the measured ADC threshold width. The intrinsic noise however does not contribute to the width. The remaining width of the threshold calculated using the data after common noise subtraction should mainly come from three components. The first one is the intrinsic variation of the threshold due to the electronics itself which is not injected into the signal, according to the electronics data sheet of the FEE used in POLAR this should be of the order of 1\,ADC~\cite{IDEAS}. The second one is propagated from the error on the common noise calculation for each event which is estimated by calculating the mean ADC value of non-triggered channels, excluding those adjacent to triggered channels, as discussed in Section~\ref{sec:pedestal_noise}. The third one is also possibly from the gain nonlinearity correction as discussed in Section~\ref{sec:non_linear}. As the maxADC used in the gain nonlinearity correction for one channel can be anyone of the other 63 channels. However for the same maxADC but of different other channels the gain nonlinearity correction factor for one channel can be slightly different because of the slight difference in thresholds of other channels. As a result there is a small error on the gain nonlinearity corrected ADC value of all events which will propagate to the calculation of the threshold width. All the three components are however not large as their sum is around 10\,ADC as shown in Figure~\ref{fig:threshold_width_noise}(b).

\subsection{Crosstalk effect}\label{sec:crosstalk}

The optical photon crosstalk exists near the PMT window between adjacent bars within one module of POLAR as mentioned in Section~\ref{sec:daq_process}. However the photon crosstalk and gain are coupled together before the energy calibration. A theoretical analysis about the the crosstalk and gain is provided in Ref.~\cite{Xiao2016}. However, as a supplement and also in order to better understand the method to calculate crosstalk matrix using in-orbit data that will be discussed in this section, a deeper understanding of the relationship between crosstalk and gain is needed. The relation between the measured ADC and the deposited energy in one module considering the optical photon crosstalk can be described by Eq.~\eqref{equ:crosstalk}.
\begin{equation}\label{equ:crosstalk}
\begin{pmatrix}
q_1 \\ q_2 \\ \vdots \\ q_{64}
\end{pmatrix}
=
\begin{pmatrix}
1 &
\frac{g_{1} m_{1,2}}{g_{2} m_{2,2}} &
\cdots & \frac{g_{1} m_{1,64}}{g_{64} m_{64,64}} \\
\frac{g_{2} m_{2,1}}{g_{1} m_{1,1}} &
1 &
\cdots &
\frac{g_{2} m_{2,64}}{g_{64} m_{64,64}} \\
\vdots & \vdots & \ddots & \vdots \\
\frac{g_{64} m_{64,1}}{g_{1} m_{1,1}} &
\frac{g_{64} m_{64,2}}{g_{2} m_{2,2}} &
\cdots & 1 \\
\end{pmatrix}
\begin{pmatrix}
g_{1} m_{1,1}\Delta E_1 \\ g_{2} m_{2,2}\Delta E_2 \\ \vdots \\ g_{64} m_{64,64}\Delta E_{64}
\end{pmatrix}
\end{equation}
where $g_i$ is the total gain of the PMT and the electronics for channel $i$, for simplicity $g_i$ also includes the light yield factor and photon collection efficiency of the PS bar which can be assumed to be the same for all bars. $[m_{ij}]$ is the photon crosstalk matrix for the module, which satisfies $\sum_{j=1}^{64} m_{ij}=1$ if assuming there is no photon leak. $q_i$ is the measured ADC value in channel $i$ which can be defined by a vector $\bm{Q}$, $\Delta E_i$ is the deposited energy in the corresponding bar which can be defined by a vector $\bm{\Delta E}$. Therefore, this equation can also be defined by Eq.~\eqref{equ:adc_energy}.
\begin{equation}\label{equ:adc_energy}
\bm{Q} = \bm{X} \cdot \bm{G} \cdot \bm{\Delta E} = \bm{X} \cdot \bm{Q'}
\end{equation}
where $\bm{X}$ is the crosstalk matrix at the ADC level with 1 for its diagonal elements, $\bm{G}$ is a diagonal matrix whose diagonal element is the reduced gain of each channel, which is $g_i m_{i,i}$, considering the coupling between the true gain and the photon crosstalk, and $\bm{Q'} = \bm{G}\cdot\bm{\Delta E} = \bm{X}^{-1}\bm{Q}$ is the vector of ADC values after the crosstalk correction.

The element of $\bm{X}$, that is $X_{ij} = \frac{g_i m_{ij}}{g_j m_{jj}}$, is actually the ratio of the signal of channel $i$ over the signal of channel $j$, that is $\frac{q_i}{q_j}$, due to crosstalk when only the channel $j$ has an energy deposition, that is $\Delta E_i = 0\; \text{where}\; i \neq j$. The element $X_{ij}$ is called the crosstalk factor from channel $j$ to channel $i$, and as it is at the ADC level it can be measured from data before energy calibration. After $\bm{X}$ is measured, the crosstalk correction can be performed by equation $\bm{Q'} = \bm{X}^{-1}\bm{Q}$.

Before doing the crosstalk correction, the measured ADC value of channel $i$ is actually $q_i = \sum_j g_i m_{ij} \Delta E_j = g_i m_{ii} \Delta E_i + \sum_{j\neq i} g_i m_{ij} \Delta E_j$, which means it not only depends on the energy deposition of the channel $i$ itself, but also depends on the energy deposition of other channels due to crosstalk. Therefore the energy calibration should be performed using the data after the crosstalk correction, that is $\bm{Q'}$. The $\bm{Q'}$ is the vector where the signal induced in one channel by the energy depositions in other channels, which is $\sum_{j\neq i} g_i m_{ij} \Delta E_j$, has been subtracted, then the elements of $\bm{Q'}$ are independent from each other. The measured gain $G_i = g_i m_{ii}$ for energy calibration of channel $i$ is then less than the true gain $g_i$, because $m_{ii}$ is less than 1, which is actually the remaining percentage of the total photons in the corresponding bar after the photon crosstalk process. It should be noted here that this was not pointed out in Ref.~\cite{Xiao2016}, in which the $\bm{Q}$ was used to perform energy calibration which is theoretically not correct from the above discussion.

\begin{figure}[!ht]
\subfigure[]{\includegraphics[height=5cm]{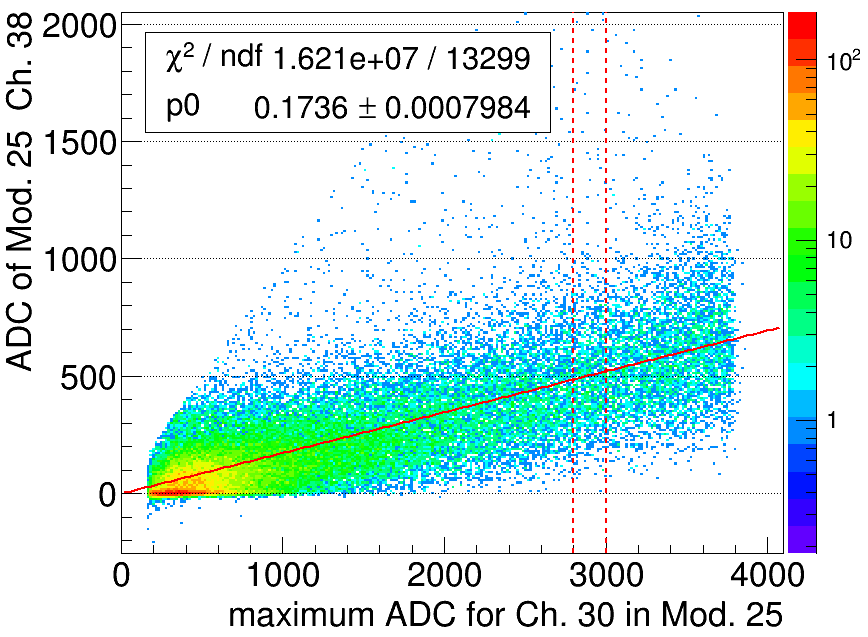}}
\hspace{2mm}
\subfigure[]{\includegraphics[height=5cm]{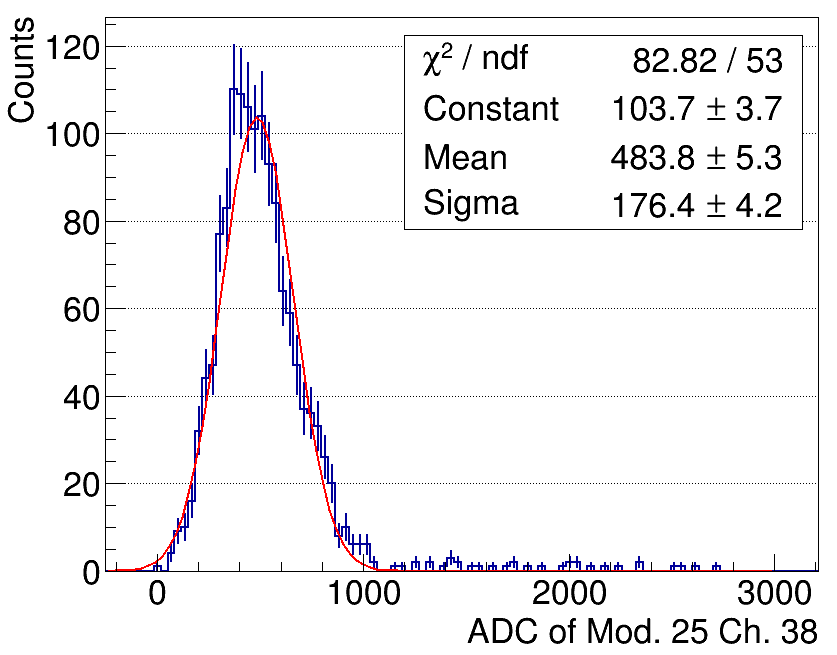}}
\caption{Crosstalk factor calculation for two adjacent channels in module 25. (a) is the correlation between the ADC values of channel 38 and channel 30, which are adjacent, where the ADC value of channel 30 at the X-axis is the maximum one in the module. (b) is the distribution of the ADC values of channel 38 when the ADC values of channel 30, as the maximum ADC in the module, are in the range 2800--3000\,ADC.}\label{fig:xtalk_calc_method}
\end{figure}

\begin{figure}[!ht]
\centering
\includegraphics[width=8cm]{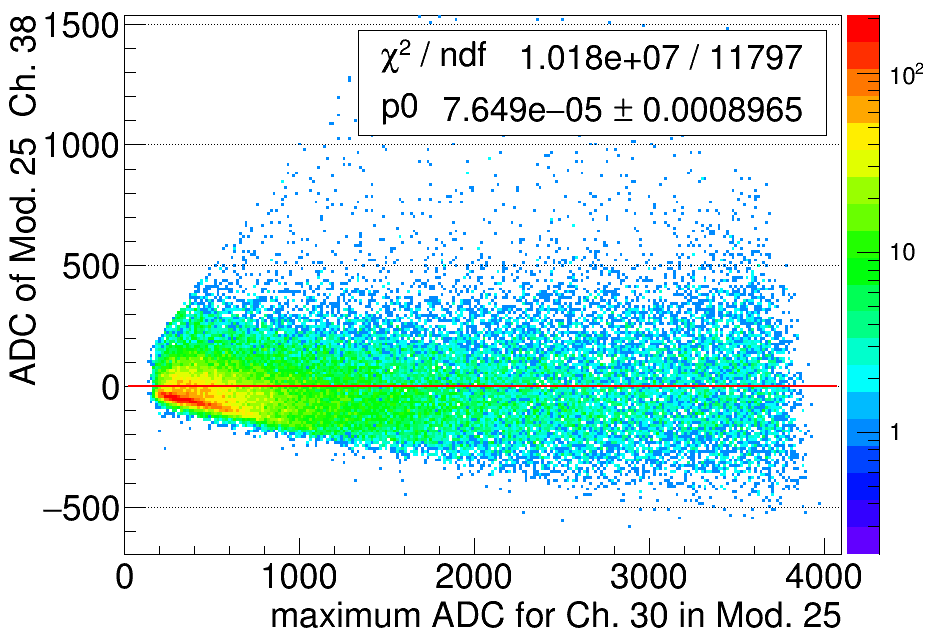}
\caption{Correlation between the ADC values of channel 38 and channel 30 of module 25 after the crosstalk correction.}\label{fig:xtalk_after_corr}
\end{figure}

To perfectly measure the crosstalk matrix $\bm{X}=[X_{ij}]$ from data at the ADC level only those events which have only one channel $j$ that has energy deposition should be selected. However this is not possible because the real energy deposition of each bar is unknown before energy calibration. But in order to calculate the crosstalk factor from channel $j$ to channel $i$, that is $X_{ij}$, an approximate method can be applied using the in-orbit data. In fact, a clear linear correlation of the ADC values between two adjacent channels due to the crosstalk effect can be found by a proper data selecting cut. Figure~\ref{fig:xtalk_calc_method}(a) shows an example of the correlation of the ADC values between two adjacent channels $i$ and $j$ for events which are selected using the cut that:

\begin{itemize}
\item channel $j$ is triggered
\item the ADC value of channel $j$, which is at the X-axis in the figure, is the maximum one in the module
\item the ADC value of channel $j$ is not in overflow
\end{itemize}

\begin{figure}[!ht]
\centering
\subfigure[]{\includegraphics[height=4.5cm]{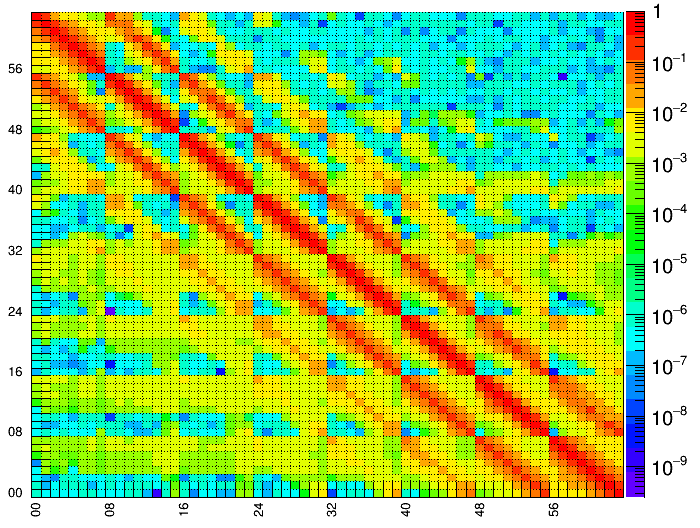}}
\hspace{2mm}
\subfigure[]{\includegraphics[height=4.5cm]{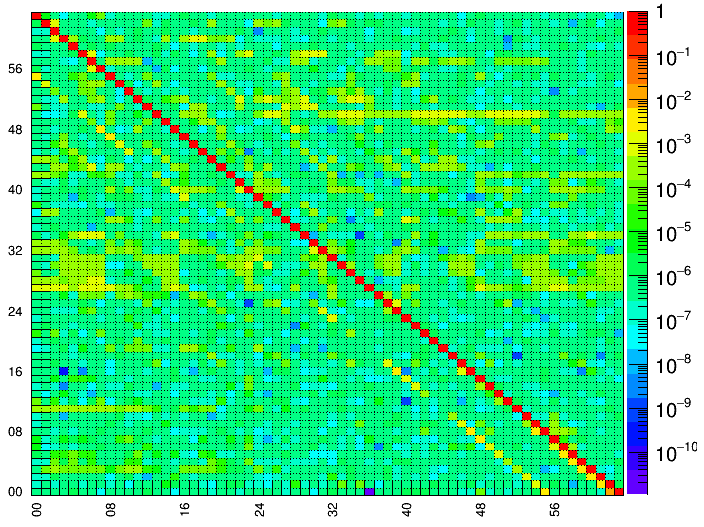}}
\caption{Crosstalk matrix of module 25, where the value of Z-axis is the crosstalk factor between two channels in log scale. (a) and (b) are respectively the crosstalk matrices of the module calculated using the data before and after crosstalk correction.}\label{fig:crosstalk_matrix}
\end{figure}

\begin{figure}[!ht]
\subfigure[]{\includegraphics[height=4.5cm]{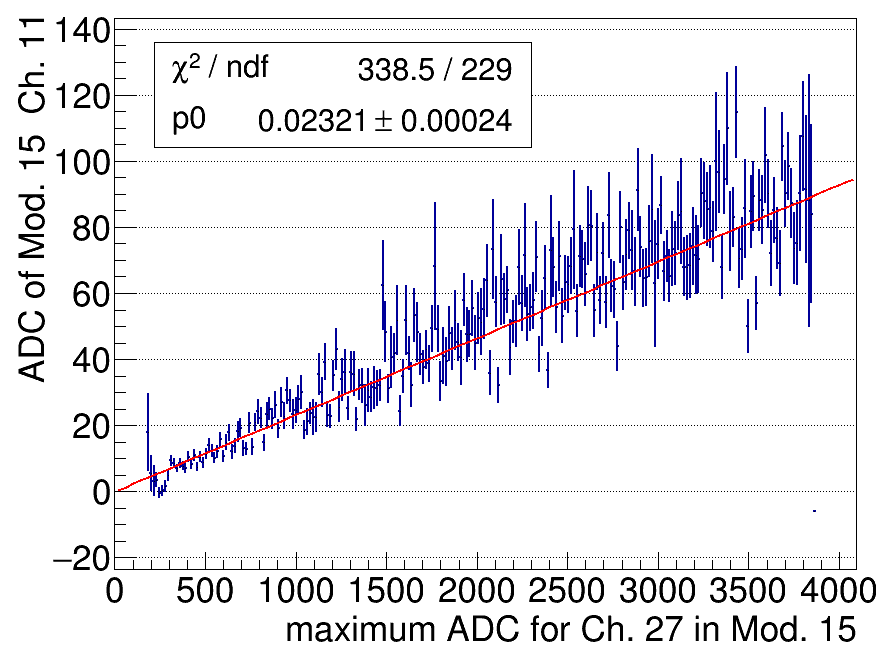}}
\hspace{2mm}
\subfigure[]{\includegraphics[height=4.5cm]{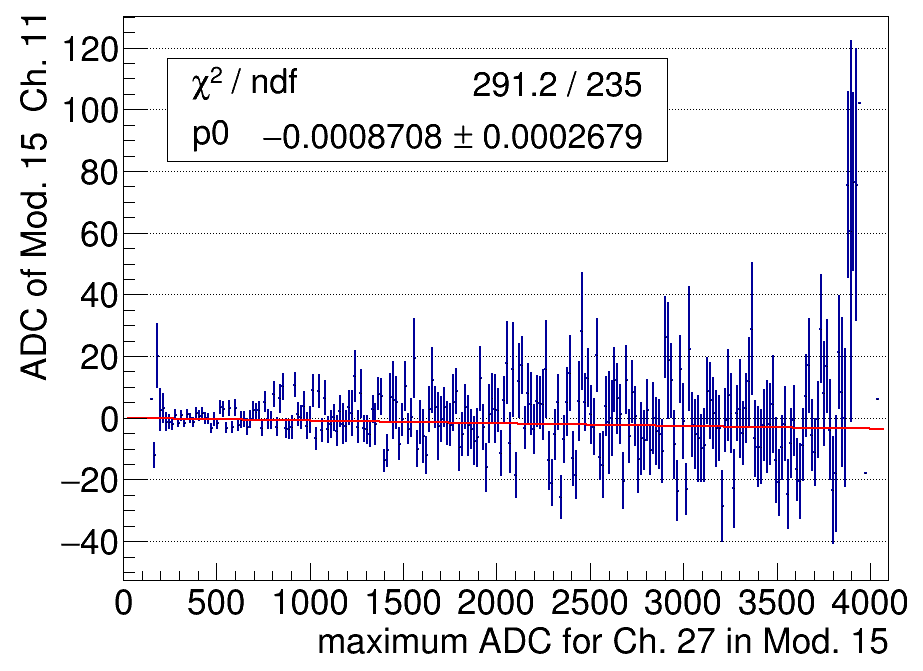}}
\caption{Crosstalk factor from channel 27 to channel 11 which is the second neighbor channel of channel 27 in module 15. (a) and (b) are respectively the crosstalk factor fitting for the two channels using the data before and after crosstalk correction.}\label{fig:xtalk_second_neighbor}
\end{figure}

As the ADC value of channel $j$ is the maximum one in the module, the great majority of the signal of channel $j$ should come from the energy deposition of the channel $j$ itself. The signal of the other one channel $i$ in the same event can come from three cases. The first case is only from the energy deposition of channel $i$ itself. The second case is only from the crosstalk signal which is mainly from channel $j$. And the third case is from the sum of the both. However compared with the crosstalk, the possibility of the case where the majority of the signal of channel $i$ is from the energy deposition of the channel $i$ itself is actually negligible, especially for those channels which are adjacent to channel $j$ which has the maximum ADC, because the crosstalk effect from channel $j$ to channel $i$ is always present whereas the fraction of the events with energy depositions in both the two channels is much smaller than 1. This can be seen from Figure~\ref{fig:xtalk_calc_method}(b) which is an example of the ADC distribution of channel $i$,  which is adjacent to channel $j$ whose ADC value is the maximum ADC in the module and is in a narrow range. The vast majority of the counts in the Gaussian distribution should come from the crosstalk signal from channel $j$ in the narrow ADC range, and the small amount of events with ADC values above the Gaussian distribution are those with the real energy deposition plus the crosstalk signal. The influence of these events with real energy deposition on the total distribution of the crosstalk signal of channel $i$ gotten from channel $j$ was found to be negligible. Each one of the elements of the crosstalk matrix $X_{ij}$, that is the crosstalk factor from channel $j$ to channel $i$, can therefore be measured by directly fitting the linear correlation shown in Figure~\ref{fig:xtalk_calc_method}(a) by function $y = k x$, and the slope parameter $k$ is an element of $X_{ij}$. Figure~\ref{fig:crosstalk_matrix}(a) shows an example of the crosstalk matrix of a full module measured using this method. Typically, the crosstalk factor from one channel to its 4 direct neighbor channels at the edge direction is of the order of $20\%$, and $5\%$ for the other 4 direct neighbor channels at the corner direction. It should be noted here that the crosstalk factor from one channel to its second neighbor channels is also significant and of the order of $2\%$. Figure~\ref{fig:xtalk_second_neighbor}(a) shows an example for this case, where the profile at the X-axis of the 2-D histogram as shown in Figure~\ref{fig:xtalk_calc_method}(a) is used instead for clarity. After the crosstalk matrix $\bm{X}$ is measured, it can then be used to correct the crosstalk effect for each event by equation $\bm{Q'} = \bm{X}^{-1}\bm{Q}$. Figure~\ref{fig:xtalk_after_corr} shows the correlation of the ADC values between the same two channels as shown in Figure~\ref{fig:xtalk_calc_method}(a) after the crosstalk correction, from which it can be seen that the correlation between the two channels is highly reduced after the crosstalk correction. And Figure~\ref{fig:xtalk_second_neighbor}(b) is the same for the case of the crosstalk from one channel to its second neighbor channels. Figure~\ref{fig:crosstalk_matrix}(b) is the crosstalk matrix of the same module as shown in Figure~\ref{fig:crosstalk_matrix}(a) but calculated using the data after the crosstalk correction, from which it can be seen that the residual crosstalk factor is very small, averagely of the order of $0.1\%$ for the adjacent channels.

\newpage

\subsection{Energy calibration}\label{sec:energy_calibration}

\begin{figure}[!ht]
\centering
\includegraphics[width=8cm]{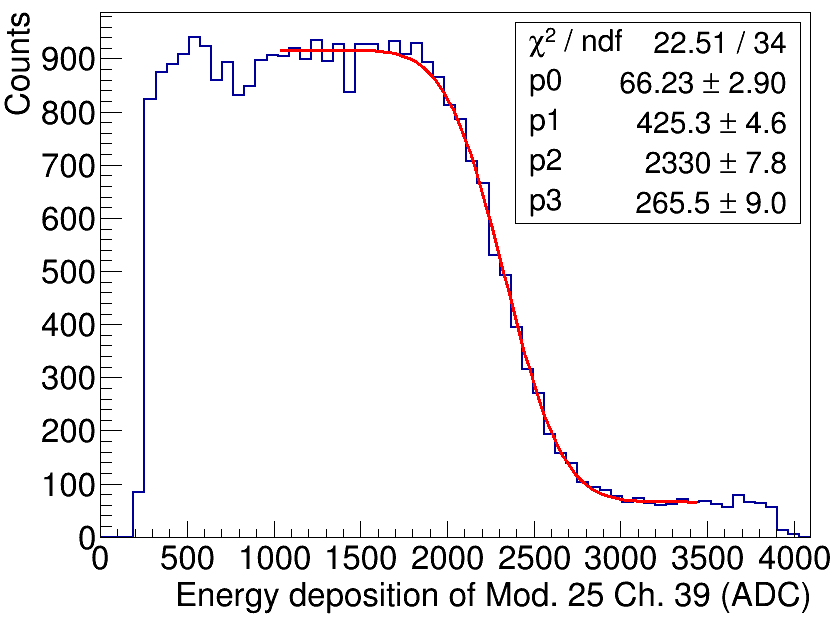}
\hspace{1mm}
\caption{The ADC spectrum induced by the 511\,keV photons from $^{22}\mathrm{Na}$ sources as measured by module 25 channel 39, the Compton Edge is centered at 340\,keV.}\label{fig:CE}
\end{figure}

The POLAR flight model contains four $^{22}\mathrm{Na}$ sources, the activity of each was approximately $100\,\mathrm{Bq}$ at the time of launch. The four sources are positioned in different corners of four different modules in the instrument and are used for the gain calibration of all 1600 channels \cite{Xiao2017}. The events of the two characteristic back-to-back emission 511\,keV photons generated by the positrons released from the $^{22}\mathrm{Na}$ sources can be selected out from the in-orbit background events by applying some geometrical cuts in combination with the knowledge of the locations of the four sources in the 2D detection plane of the instrument. As discussed in Section~\ref{sec:crosstalk}, the energy calibration work should be performed based on the data after crosstalk correction. With the geometrical cuts optimized to select out those events induced by the $^{22}\mathrm{Na}$ sources as much as possible and keep the contamination from in-orbit background events at a minimum, approximately 12 hours of the in-orbit data suffices to produce a spectrum with a clear Compton Edge (CE) of the 511\,keV photons induced by the sources for each channel, despite the high radiation environment in which the data is taken. With the 12 hours of in-orbit data, the relative measurement error of CE for more than $90\%$ channels can be less than $3\%$. An example of the spectrum in ADC for one channel generated using about 24 hours of in-orbit data is shown in Figure~\ref{fig:CE}. This spectrum can be fitted by the function presented in Eq~\eqref{equ:CE}.

\begin{figure}[!ht]
\centering
\subfigure[]{\includegraphics[height=5.0cm]{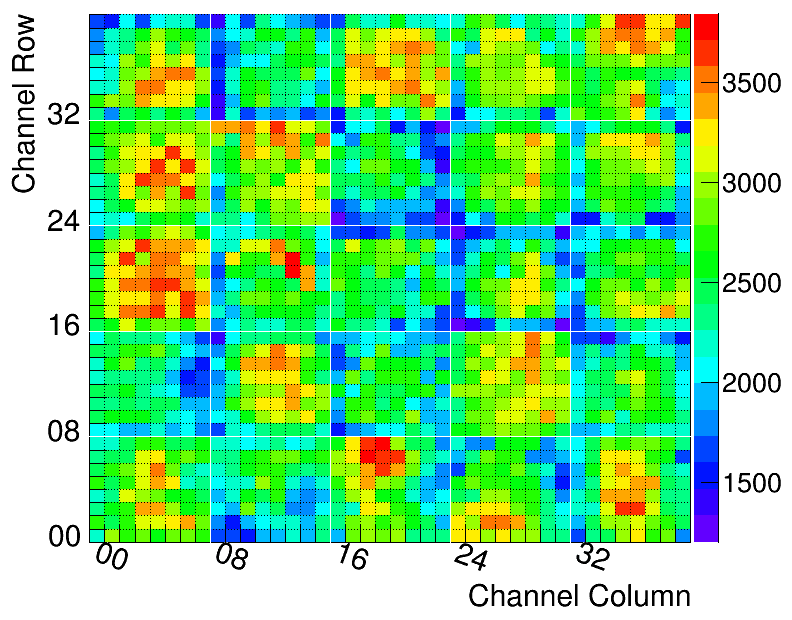}}
\subfigure[]{\includegraphics[height=5.0cm]{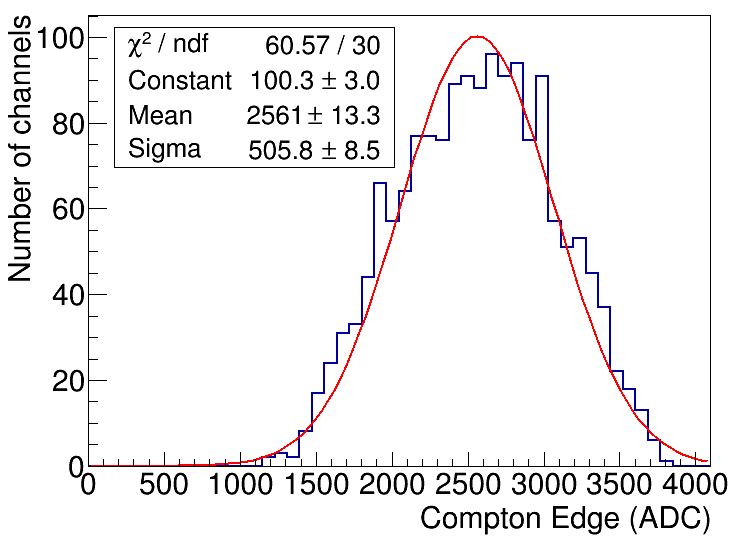}}
\caption{Compton Edge positions of all 1600 channel corresponding to the 511 keV photons as measured in ADC using the in-orbit data acquired with a specific high voltage setting for energy calibration. (a) shows the map and (b) shows the distribution.}\label{fig:CE_pos}
\end{figure}

\begin{equation}\label{equ:CE}
f(x) = p_0 + p_1 \times \mathrm{erfc}\left(\frac{x - p_2}{\sqrt{2}p_3}\right)
\end{equation}
where $p_2$ is the position of CE, $p_3$ is the width of CE, $p_1$ and $p_0$ are two free parameters considering the quantity of statistics and the contamination from in-orbit background events. Figure~\ref{fig:CE_pos} shows the map and distribution of the CE positions, corresponding to the 511\,keV photons, of all 1600 channels measured by this method from in-orbit data in a specific high voltage setting for energy calibration. As it is known that the energy corresponding to the CE of 511\,keV photons is approximately 340\,keV, the gain of each channel, with unit ADC/keV, can be acquired by dividing the CE position of the channel by 340\,keV.

\newpage

\section{Temperature and high voltage dependence}\label{sec:temp_hv_dependence}

\begin{figure}[!ht]
\centering
\includegraphics[width=11cm]{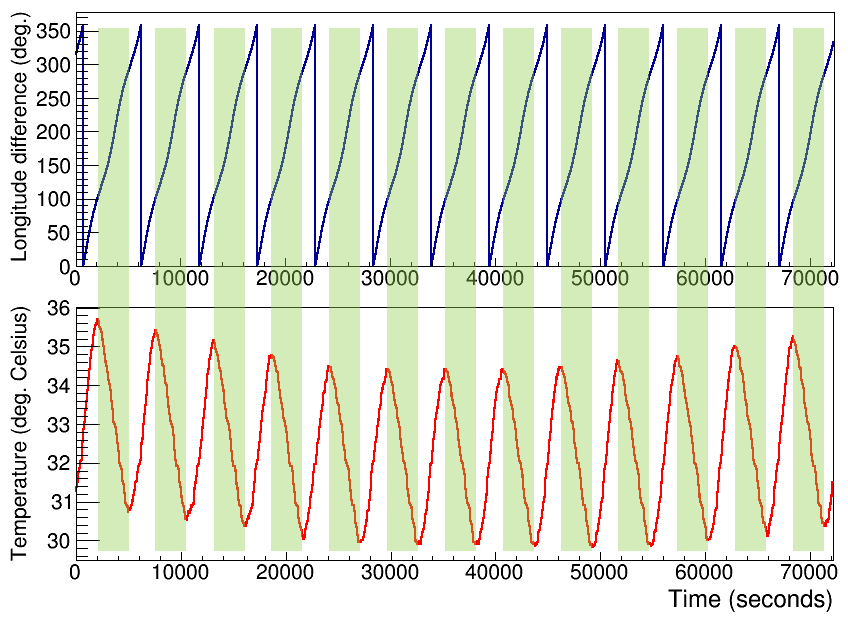}
\caption{Changing of the average temperature of 25 FEEs (bottom) together with the longitude difference between the positions of TG-2 and the Sun (top) in about 20 hours, the green (or gray) superimposed box illustrates the shadow of the Earth, where a longitude difference of $0^\circ$ corresponds to TG-2 being right in between the Sun and the Earth, $90^\circ$ corresponds to TG-2 starting to go into the shadow while $270^\circ$ corresponds to TG-2 starting to move out the shadow.}\label{fig:temperature_change}
\end{figure}

The in-orbit environment is changing with time, the most important of which considering the effect on the calibration parameters that are discussed in Section~\ref{sec:calib_pars} is the temperature of the instrument. Figure~\ref{fig:temperature_change} shows the change of the average temperature of 25 FEEs in about 20 hours (bottom), together with the longitude difference between the positions of TG-2 and the Sun (top). It can be seen that the average temperature of 25 FEEs goes down and up continuously in different orbits when TG-2 is in and out of the shadow of the Earth while a maximum temperature is reached right before TG-2 moves into the shadow of the Earth. The change of the average temperature in one orbit is about $5^\circ\mathrm{C}$. Considering some other effects like the attitude change of TG-2, the biggest change of the average temperature in a long period is within about $10^\circ\mathrm{C}$. Therefore the dependence on temperature of those calibration parameters are studied in Sections from \ref{sec:ped_temp} to \ref{sec:gain_temp}.

In order to reduce the energy threshold as much as possible to be able to increase the detection sensitivity, the high voltage setting of the PMTs for the normal scientific data acquisition are relatively high. As a result the Compton Edges of the 511\,keV photons from the $^{22}\mathrm{Na}$ sources for most channels are in overflow, that is larger than 4095\,ADC. Therefore it cannot be measured directly from the in-orbit data taken with the normal data acquisition high voltage. As it is known that there is a linear relationship between the gain and high voltage setting in log scale, the gain parameter of each channel for the normal data acquisition high voltage can therefore be calculated using the linear relationship that can be measured from the data taken with several different lower high voltage settings. The dependence on high voltage of the gain parameter has already been measured using the ground data before launch for the purpose of optimizing the in-orbit high voltage setting~\cite{Zhang2018}. However the measurement of this dependence should be redone using the in-orbit data in case there are some changes due to, for example, vibrations and shocks sustained by the PMTs during launch. The measurement of the relationship between gain and high voltage using in-orbit data and the final results for the gain in the normal operating condition are discussed in Section~\ref{sec:gain_hv}.

\subsection{Pedestal vs. temperature}\label{sec:ped_temp}

\begin{figure}[!ht]
\centering
\subfigure[]{\includegraphics[height=5cm]{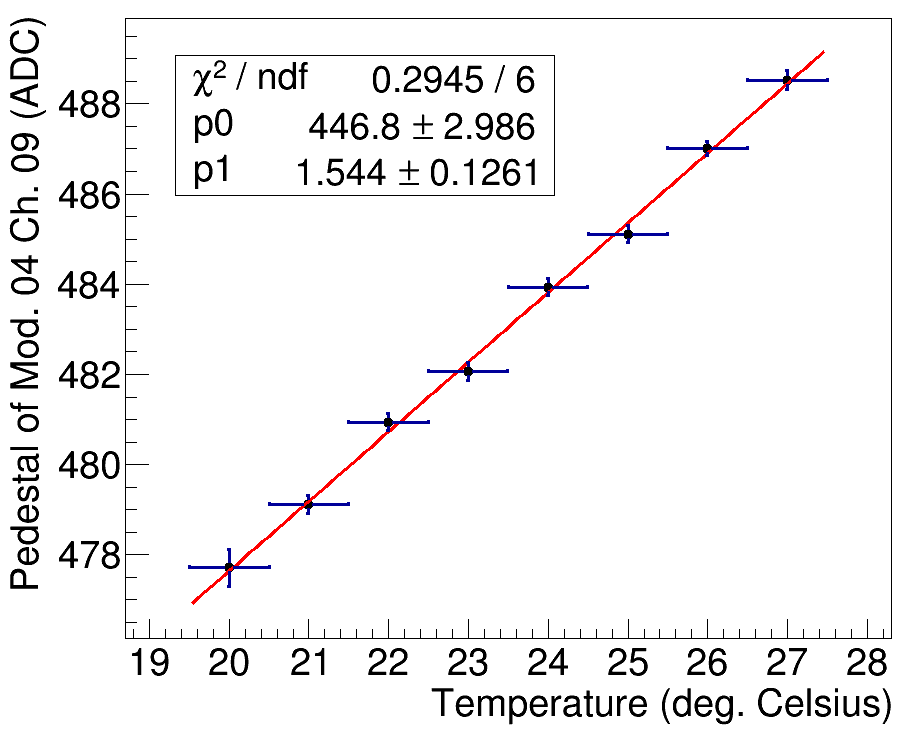}}
\hspace{2mm}
\subfigure[]{\includegraphics[height=5cm]{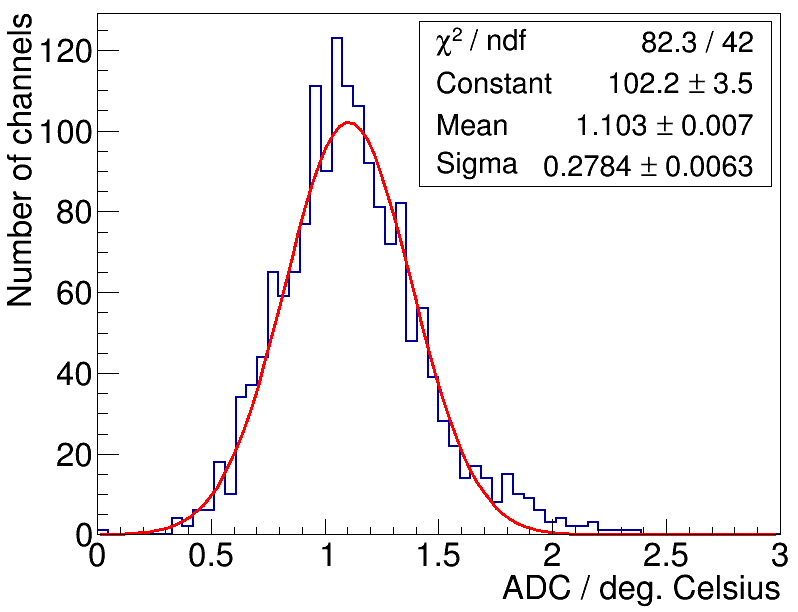}}
\caption{Dependence of pedestal level on temperature. (a) is the correlation between pedestal level and temperature and the linear fitting for module 04 channel 09. (b) is the distribution of the slope of the pedestal level change with temperature for all 1600 channels.}\label{fig:ped_vs_temp}
\end{figure}

The pedestal levels of all channels were found to have a small but non-negligible dependence on temperature. The dependence of the pedestal level on temperature of each channel was measured by firstly binning the pedestal ADC values of the channel as measured using the pedestal events in different temperature bins then fitting the correlation by a linear function. The temperature was measured by the temperature sensor mounted on each FEE, whose precision is $1\,^\circ\mathrm{C}$. One example of the correlation between the pedestal level and temperature together with a linear fitting for one channel is shown in Figure~\ref{fig:ped_vs_temp}(a). The points in the figure were measured using the in-orbit data of about 4 continuous days, and a clear linear correlation was found for all channels in the temperature range of the in-orbit environment of POLAR. Figure~\ref{fig:ped_vs_temp}(b) shows the distribution of the slope values acquired from the linear fitting on the correlation of all channels. It can be seen that the average slope is about $1\,\mathrm{ADC}/^\circ\mathrm{C}$. As the temperature of all modules is continuously changing with time, the pedestal level of each channel at different temperatures can be calculated by the linear correlation after it is measured, then subtracted from physical events in the data processing pipeline.

\subsection{Common noise and intrinsic noise vs. temperature}

With the same 4 days of in-orbit data as used in the study of the dependence of pedestal level on temperature, the dependence of common noise and intrinsic noise on temperature were also studied. The common noise of each module and the intrinsic noise of each channel are calculated using the method discussed in Section~\ref{sec:pedestal_noise}, but using only pedestal events as physical events will contaminate the common noise measurement. Subsequently the correlation between the width of the common noise (and intrinsic noise) and the temperature of the module was measured. The width of the common noise and the intrinsic noise for each temperature are acquired by performing a fitting with a Gaussian function on the distribution of the two kinds of noise measured using the sample of pedestal events with the same temperature. 

\begin{figure}[!ht]
\centering
\subfigure[]{\includegraphics[height=5cm]{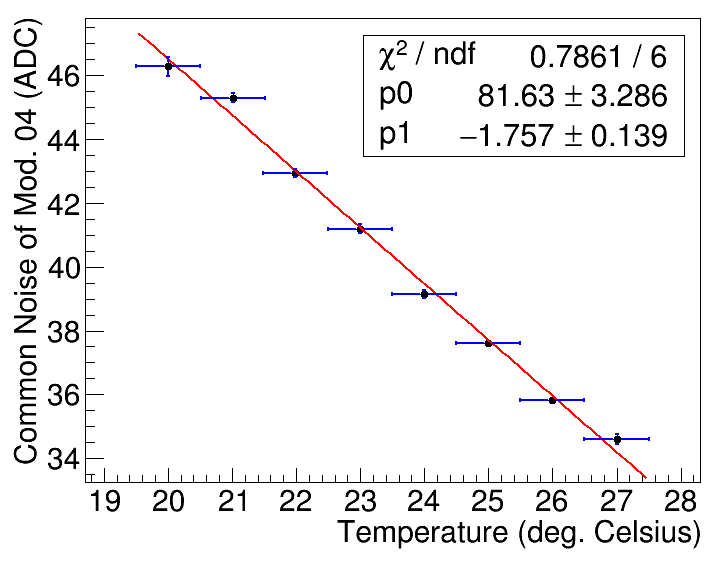}}
\hspace{2mm}
\subfigure[]{\includegraphics[height=5cm]{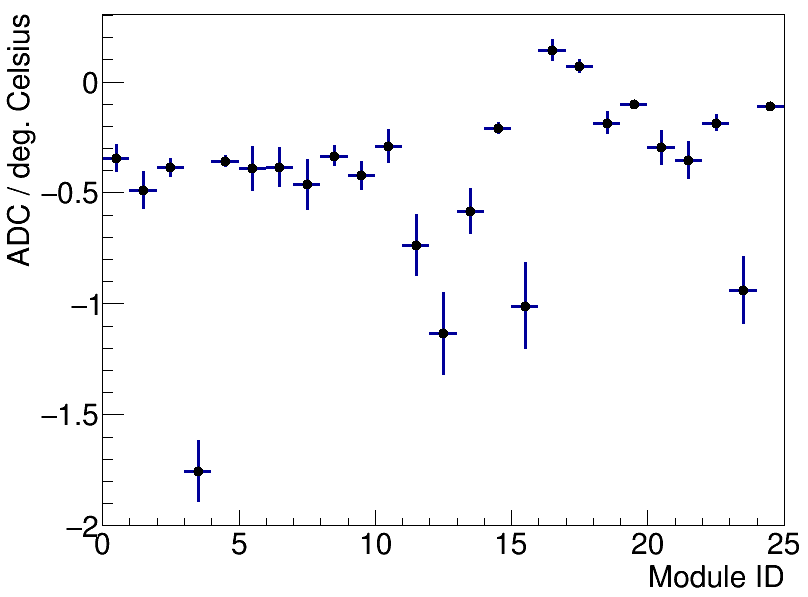}}
\caption{Dependence of common noise width on temperature. (a) is the correlation between common noise width and temperature and the linear fitting for module 04. (b) is the slope of the common noise width change with temperature for all 25 modules}\label{fig:common_noise_vs_temp}
\end{figure}

\begin{figure}[!ht]
\centering
\subfigure[]{\includegraphics[height=4.8cm]{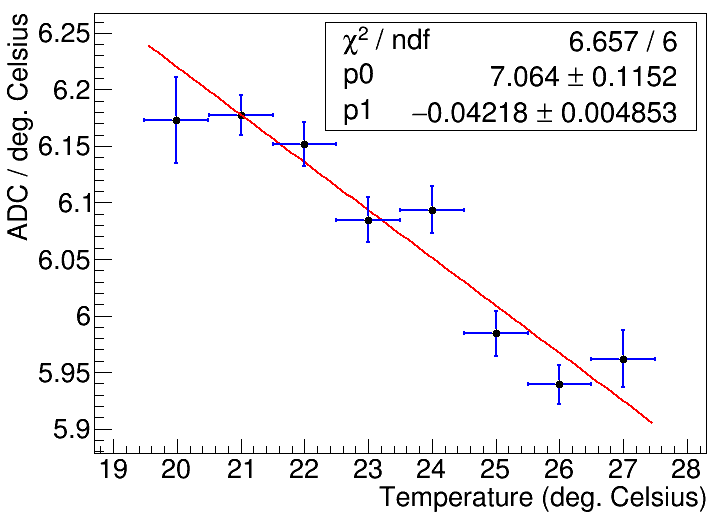}}
\hspace{2mm}
\subfigure[]{\includegraphics[height=4.8cm]{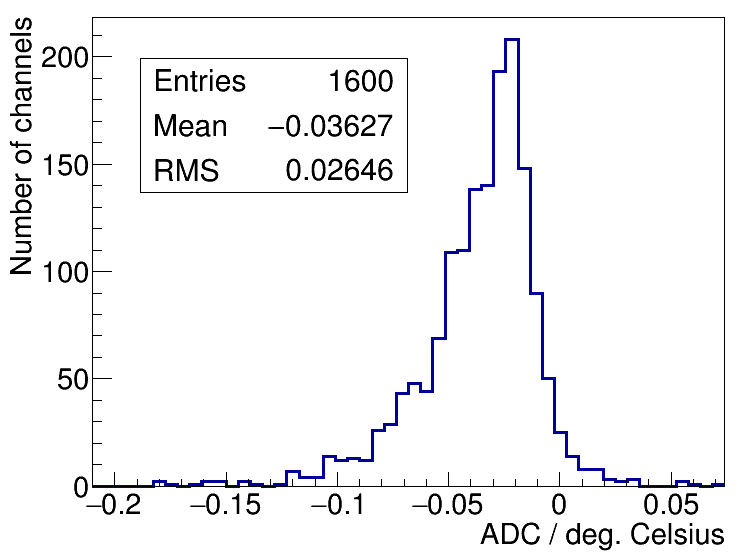}}
\caption{Dependence of intrinsic noise on temperature. (a) is the correlation between intrinsic noise width and temperature and the linear fitting for module 04 channel 56. (b) is the distribution of the slope of the intrinsic noise width change with temperature for all 1600 channels.}\label{fig:intrinsic_noise_vs_temp}
\end{figure}

It was found that the common noise width also has a significant dependence on temperature for most modules as shown in Figure~\ref{fig:common_noise_vs_temp}(a) and it can be fitted relatively well using a linear function. Figure~\ref{fig:common_noise_vs_temp}(b) shows the slope values acquired from fitting the common noise width as a function of temperature with a linear function for all 25 modules. It can be seen that the common noise width decreases slightly when the temperature increases for most modules. The average slope is around $-0.5\,\mathrm{ADC}/^\circ\mathrm{C}$, whereas for a few modules the linear correlation between the common noise width and temperature is not very significant, as a result the slope value of the linear fitting is close to zero. As the common noise is calculated and subtracted event by event in the data reduction pipeline of POLAR, the dependence of the common noise width on temperature does not influence the data analysis. The study is however useful for performing an accurate Monte Carlo simulation in the digitization process, where different common noise values need to be applied according to different temperatures.

The same procedure was performed for the intrinsic noise. Figure~\ref{fig:intrinsic_noise_vs_temp}(a) shows an example of the fitting using a linear function on the correlation between the intrinsic noise width and the temperature of one channel. And Figure~\ref{fig:intrinsic_noise_vs_temp}(b) shows the distribution of the slope of all 1600 channels. It can be found that the dependence on temperature for intrinsic noise is not very strong. The intrinsic noise width also slightly decreases when the temperature increases but the slope is less than $0.1\,\mathrm{ADC}/^\circ\mathrm{C}$ for most channels. This means the dependence is negligible in the temperature range of the in-orbit environment of POLAR which varies by a maximum of about $10\,^\circ\mathrm{C}$ .

\subsection{Gain nonlinearity and threshold vs. temperature}

\begin{figure}[!ht]
\centering
\subfigure[]{\includegraphics[height=4.4cm]{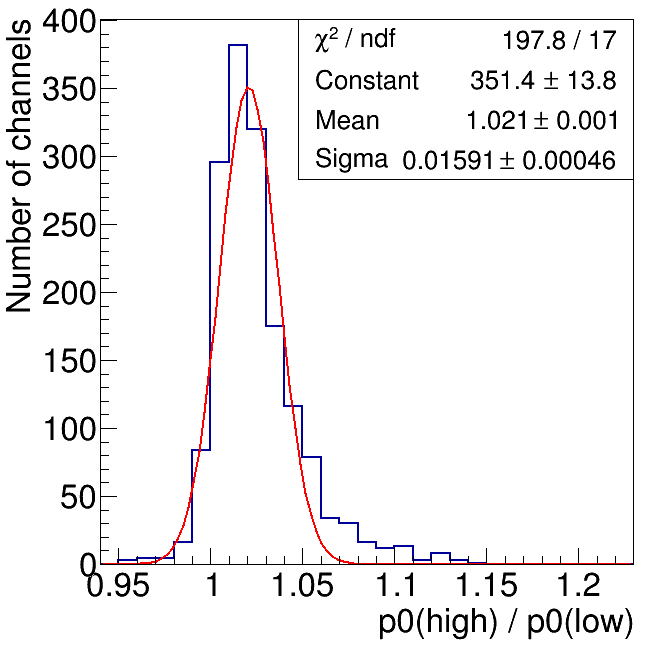}}
\subfigure[]{\includegraphics[height=4.4cm]{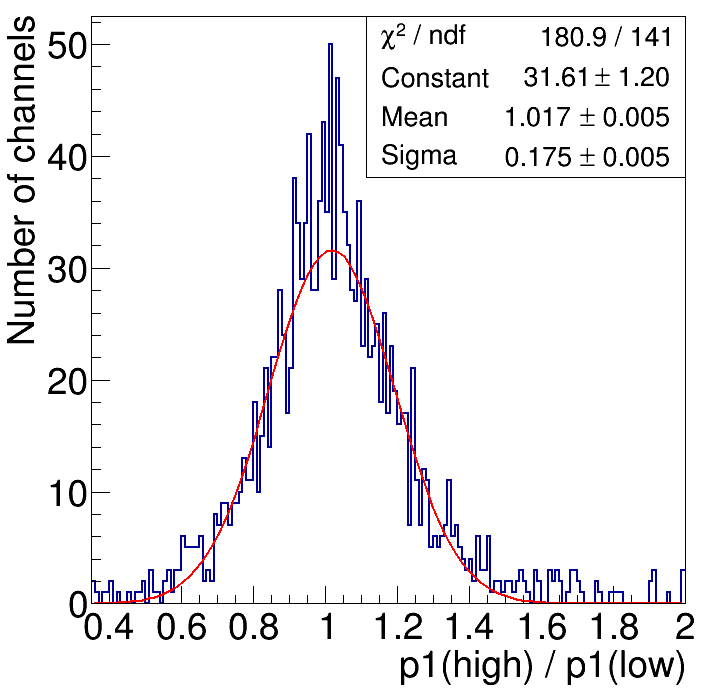}}
\subfigure[]{\includegraphics[height=4.4cm]{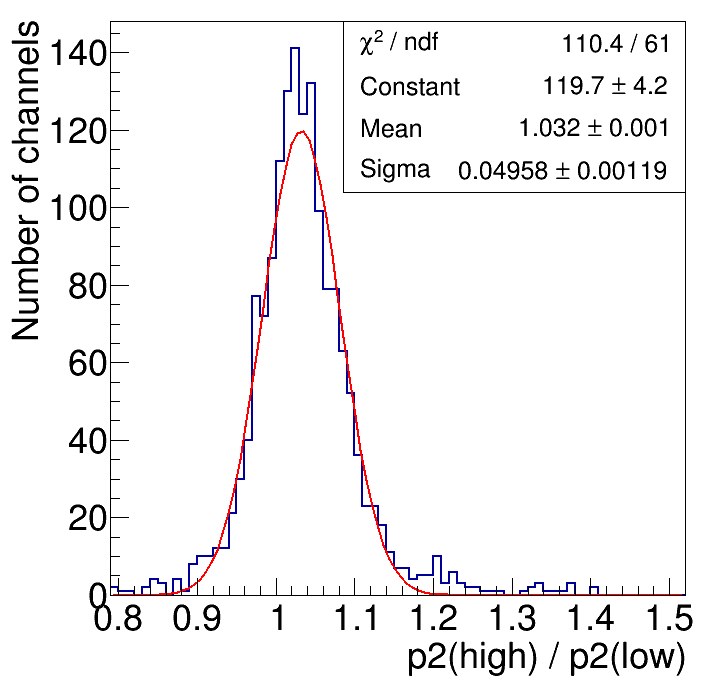}}
\caption{Distribution of the ratio of $p_n(\text{high temperature}) / p_n(\text{low temperature})$ for the comparison of the nonlinearity functions calculated using data with different temperatures. (a), (b) and (c) are respectively for the ratio of $p_0$, $p_1$ and $p_2$.}\label{fig:nonlin_temp}
\end{figure}

As discussed in Section~\ref{sec:non_linear}, the nonlinearity function of the gain of each channel as shown in Eq.~\eqref{equ:non-linear} has three parameters, which are $p_0$, $p_1$ and $p_2$. In order to study the temperature effect on the three parameters of the nonlinearity function of the gain of each channel, the nonlinearity functions that are calculated using data with two different temperatures are compared. The two different temperatures chosen for each module for the comparison are the highest and the lowest temperature of the module in several orbits. And the differences of the two temperatures for different modules are between $3^\circ\mathrm{C}$ and $6^\circ\mathrm{C}$, because the temperature range of different modules within the same time period are slightly different. The distributions of the ratio of the three parameters calculated using data with high temperature over those with low temperature are shown in Figure~\ref{fig:nonlin_temp}.

It can be seen that the mean values of the ratio of the three parameters are all slightly larger than 1, which means all the three parameters show a small but statistically significant increase when the temperature increases $3-6^\circ\mathrm{C}$. The average increase percentage of $p_0$, $p_1$ and $p_2$ are $2.1\%$, $1.7\%$ and $3.2\%$ respectively. In the gain nonlinearity correction process the normalized nonlinearity function as shown in Eq.~\eqref{equ:non-linear-corr} is applied, therefore the change of $p_0$ has no influence. Only $p_1$ and $p_2$ can affect the shape of the normalized nonlinearity function. It is found through numerical calculation that if the relative variation of $p_1$ and $p_2$ are $4\%$, the variation propagated to the final gain nonlinearity corrected ADC value of one channel is less than 5\;ADC. Therefore it is acceptable for the nonlinearity function of each channel to be calculated using all data and to be applied on all data without taking the temperature change into account.

\begin{figure}[!ht]
\centering
\subfigure[]{\includegraphics[height=5cm]{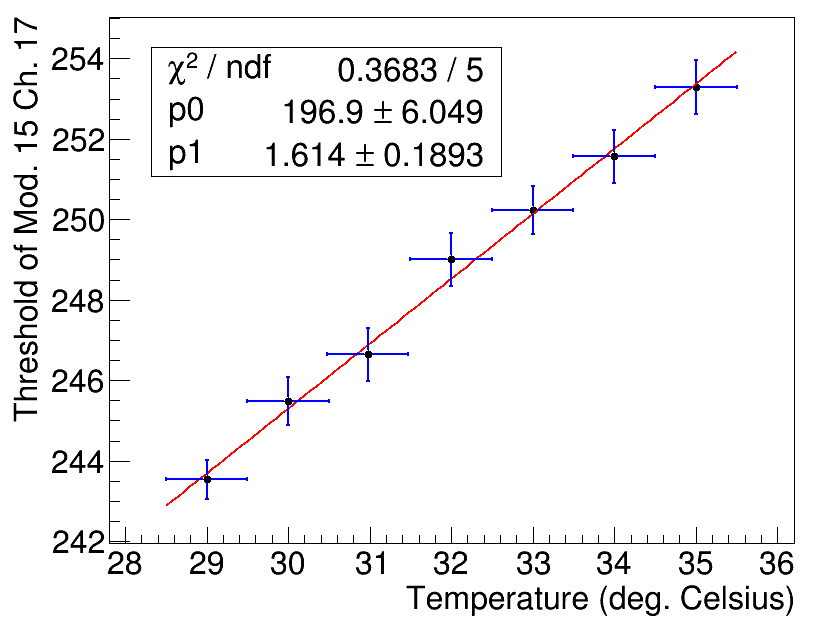}}
\hspace{2mm}
\subfigure[]{\includegraphics[height=5cm]{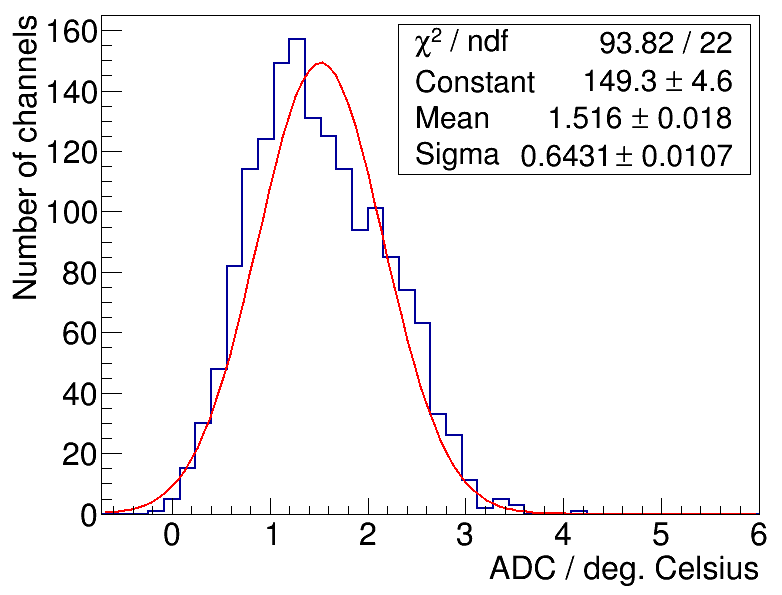}}
\caption{Dependence of ADC threshold on temperature. (a) is the correlation between ADC threshold and temperature and the linear fitting for module 15 channel 17. (b) is the distribution of the slope of the ADC threshold change with temperature for all 1600 channels.}\label{fig:vthr_temp}
\end{figure}

As the nonlinearity function of each channel is calculated by fitting the ADC threshold position as a function of maxADC using the data before gain nonlinearity correction, the parameter $p_0$ actually has a linear correlation with the ADC threshold position calculated using the data after gain nonlinearity correction, which is independent on maxADC. It can be seen from Figure~\ref{fig:nonlin_temp}(a) that $p_0$ shows a statistically significant increase with temperature, which means the ADC threshold that is calculated using the gain nonlinearity corrected data should also increase with temperature. Therefore a more detailed study of the dependence of the ADC threshold on temperature was performed. It was found that the ADC threshold has a clear linear correlation with temperature and increases with temperature for most channels, at least in the in-orbit temperature range of POLAR. Figure~\ref{fig:vthr_temp}(a) shows an example of the correlation between the ADC threshold position and temperature for one channel. The distribution of the slope of the linear fitting on the correlation between ADC threshold and temperature for all channels is shown in Figure~\ref{fig:vthr_temp}(b). It can be seen that the mean slope is about $1.5\;\mathrm{ADC}/^\circ\mathrm{C}$ for the threshold position. As this correlation is significant for most channels as shown in Figure~\ref{fig:vthr_temp}(a), the dependence of threshold on temperature should also be taken into account in the digitization process of the Monte Carlo simulation.

\subsection{Crosstalk vs. temperature}

\begin{figure}[!ht]
\centering
\includegraphics[height=6cm]{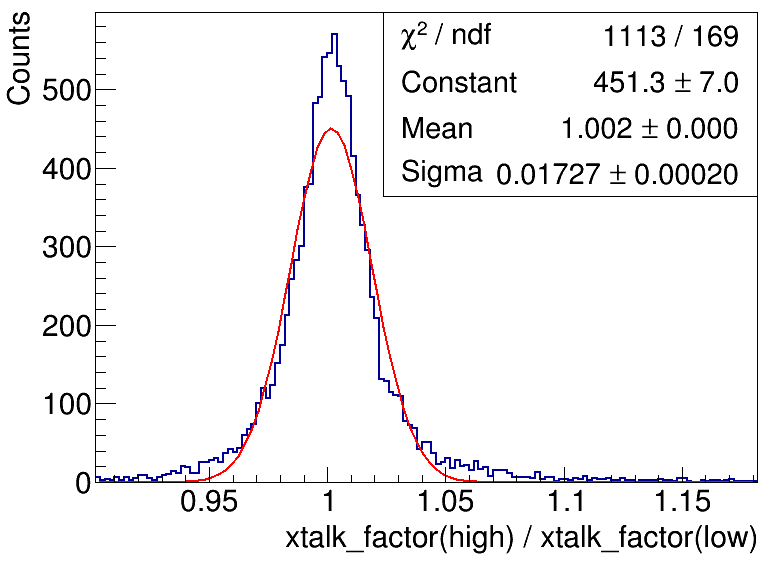}
\caption{Distribution of the ratio of crosstalk factor with high temperature over that with low temperature for directly adjacent channels.}\label{fig:xtalk_factor_temp}
\end{figure}

Similar to the check for the effect of temperature on nonlinearity function, the same check was also performed for crosstalk matrix using the same data. Two crosstalk matrices for each module were calculated using the data with the highest and the lowest temperature of the module, then the distribution of the ratio of the crosstalk factor with high temperature over that with low temperature was checked. Here only the crosstalk factors between two directly adjacent channels were checked because the crosstalk between adjacent channels dominates. The distribution of the ratio is shown in Figure~\ref{fig:xtalk_factor_temp}. It can be seen that the mean ratio is very close to 1 and the average relative change is only $0.2\%$ in the temperature range $3-6^\circ\mathrm{C}$, although the measurement of the difference is statistically significant it is negligible for the analysis of POLAR data. Therefore the temperature effect for crosstalk correction is also not necessary to be taken into account in the data reduction pipeline of POLAR.

As discussed in Section~\ref{sec:crosstalk}, the definition of the crosstalk factor from channel $j$ to channel $i$ is $X_{ij} = \frac{g_i m_{ij}}{g_j m_{jj}}$. The effect of temperature on the photon crosstalk matrix $[m_{ij}]$ should be negligible because the photon crosstalk occurs at the bottom of PS bars and has no significant relation with the electronics. If temperature can affect the crosstalk factor the effect can only come from $g_i/g_j$, which is the relative change of the true gain between channel $i$ and channel $j$ for two temperatures, and from Figure~\ref{fig:xtalk_factor_temp} it can be seen that this relative change is also negligible.

\subsection{Gain vs. temperature}\label{sec:gain_temp}

\begin{figure}[!ht]
\centering
\subfigure[]{\includegraphics[height=5cm]{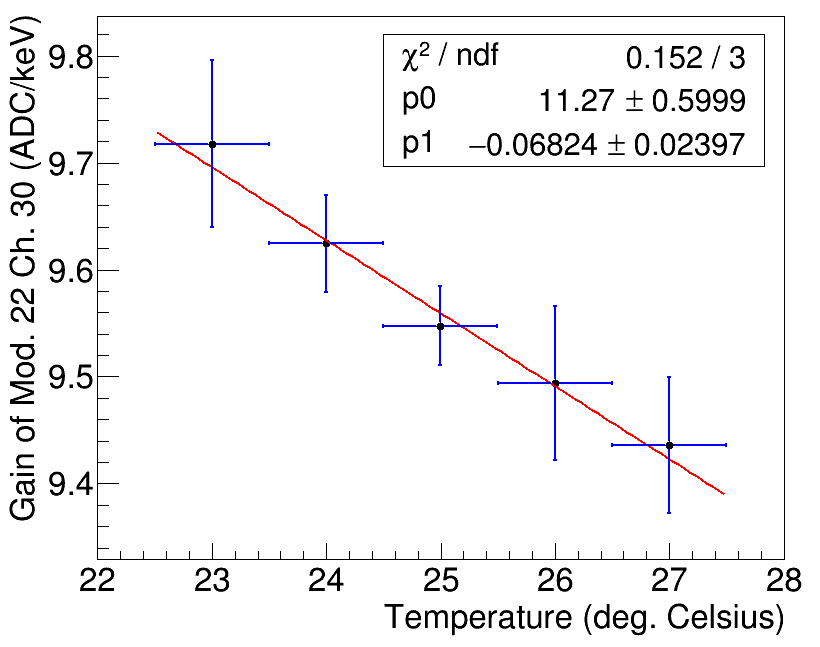}}
\hspace{2mm}
\subfigure[]{\includegraphics[height=5cm]{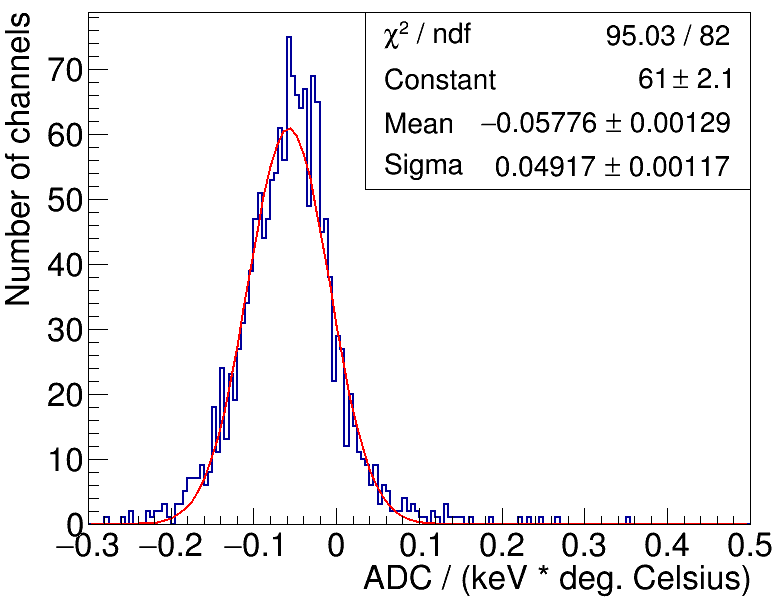}}
\caption{Dependence of gain on temperature. (a) is the correlation between gain and temperature and the linear fitting for module 22 channel 30. (b) is the distribution of the slope of the gain change with temperature for all 1600 channels.}\label{fig:gain_vs_temp}
\end{figure}

In order to study the dependence of gain on temperature, more than 5 days of in-orbit data is used, as the statistics quantity of the $^{22}$Na source events of one day of data is enough for each temperature in the temperature range of about $5^\circ\mathrm{C}$ for each module. It is found that for most channels the gain has a negative correlation with temperature. Figure~\ref{fig:gain_vs_temp}(a) shows one example of the correlation and Figure~\ref{fig:gain_vs_temp}(b) shows the distribution of the slope of the linear fitting on the correlation for all channels. And it can be seen that the average slope of the gain change with temperature is about $-0.058\;\mathrm{ADC}/(\mathrm{keV}\cdot\,^\circ\mathrm{C})$. As the average gain of the high voltage for the 5 days of data is about 8~ADC/keV, the average relative change is about $-0.7\%/^\circ\mathrm{C}$, and in the typical $5^\circ\mathrm{C}$ temperature range of POLAR the average relative change is of the order of $4\%$.

It should be noted here that there is a significant amount of channels (about half) where the correlation between the gain and temperature can hardly be described using a linear function, even when the statistics quantity is sufficient. However, the effect of the average $4\% $ relative change of gain on the final modulation curve needs to be studied by Monte Carlo simulation in order to understand the effect of not applying the temperature correction for gain on a channel by channel basis. Such a study was performed and will be discussed in Section~\ref{sec:calib_sys_error}.

\subsection{Gain vs. High Voltage}\label{sec:gain_hv}

\begin{figure}[!ht]
\centering
\includegraphics[width=8cm]{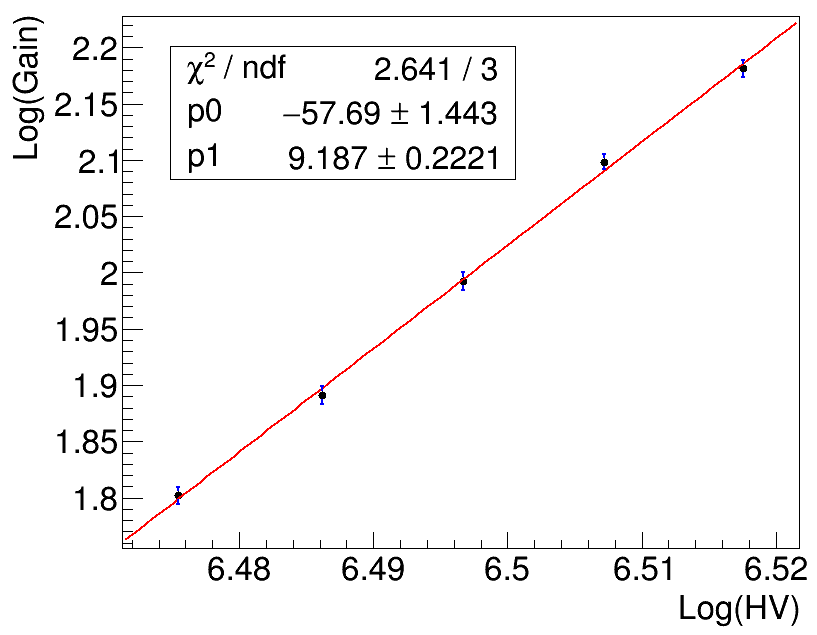}
\caption{The relationship between $\log(Gain)$ and $\log(HV)$ and the linear fitting for module 25 channel 07. Here the natural logarithm is used.} \label{fig:gain_vs_hv}
\end{figure}

\begin{figure}[!ht]
\centering
\subfigure[]{\includegraphics[height=5.2cm]{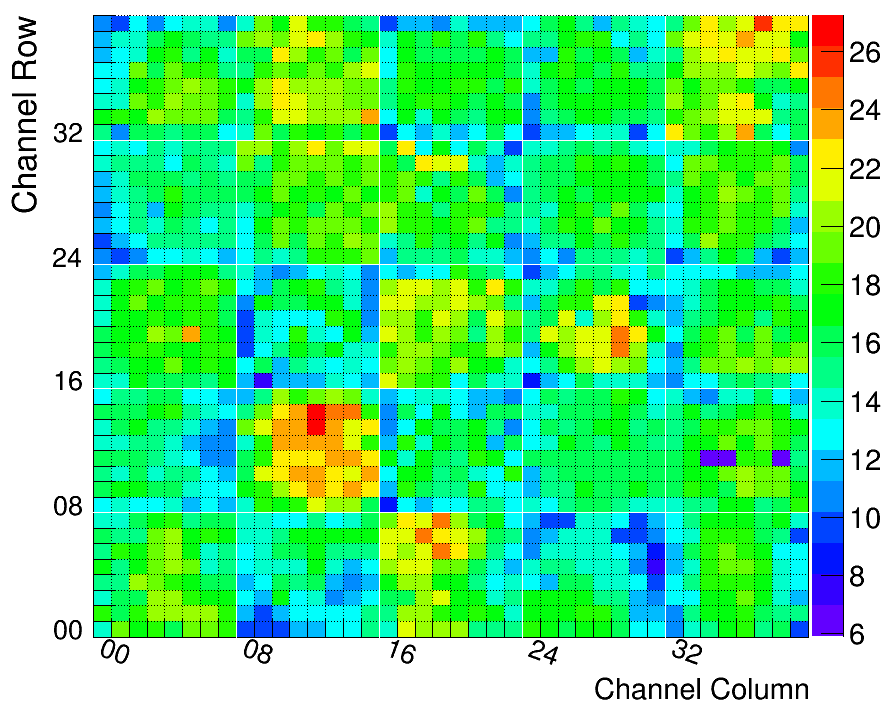}}
\hspace{2mm}
\subfigure[]{\includegraphics[height=5.2cm]{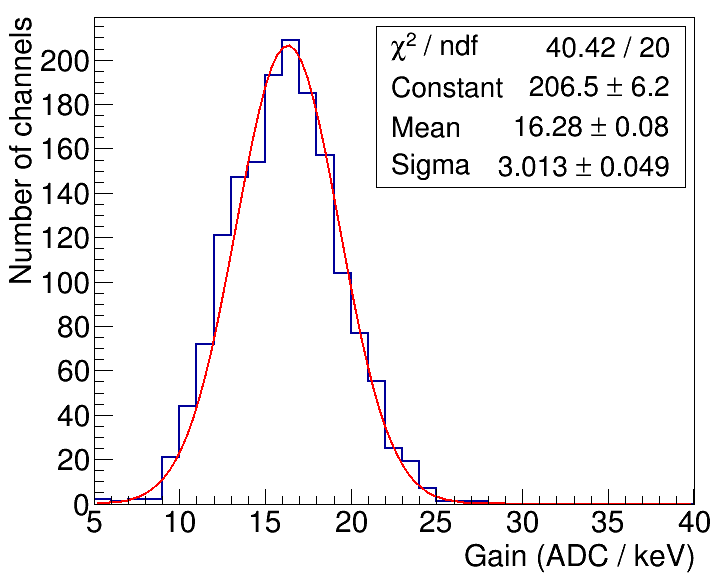}}
\caption{The final calculated gain map (a) of the normal data acquisition high voltage setting, and the corresponding distribution of the gain (b) for all 1600 channels.}\label{fig:gain_bank50}
\end{figure}

In order to measure the relationship between the gain and high voltage for each channel to calculate the gain corresponding to the high voltage setting for the normal scientific data acquisition, 5 different lower high voltage settings were applied in November 2016 and one day of in-orbit data was acquired for each of the 5 lower high voltage settings. Using the method described in Section~\ref{sec:energy_calibration}, a map of the Compton Edge position as well as the fitting error of all 1600 channels corresponding to the 511~keV photons induced by the $^{22}$Na sources can be produced for each of the 5 lower high voltage settings. 
As discussed in Section~\ref{sec:gain_temp}, the gain of one channel for the same high voltage setting can also be slightly affected by temperature, however if only using the data with the same temperature the statistics quantity of the ADC spectrum for each temperature within one day will not be sufficient. Therefore all the one day of data for each high voltage setting is used to generate the ADC spectrum from the $^{22}$Na source events, as a result, the measured Compton Edge of each channel is actually the mean one in the about $5^\circ\mathrm{C}$ temperature range within one day, but has a smaller error. As discussed in Ref.~\cite{Zhang2018}, the Gain--HV relationship for each channel can be described by Eq.~\eqref{equ:gain_vs_hv}.
\begin{equation}\label{equ:gain_vs_hv}
\log(Gain) = p_0 + p_1 \log(HV)
\end{equation}
This relation is well confirmed by the analysis of in-orbit data as shown for one channel in Figure~\ref{fig:gain_vs_hv}. And using this Gain--HV relationship measured by in-orbit data for each channel the final calculated gain map as well as the distribution, with unit ADC/keV, of the high voltage setting used for the normal scientific data acquisition is shown in Figure~\ref{fig:gain_bank50}.

\newpage

\begin{figure}[!ht]
\centering
\includegraphics[width=8cm]{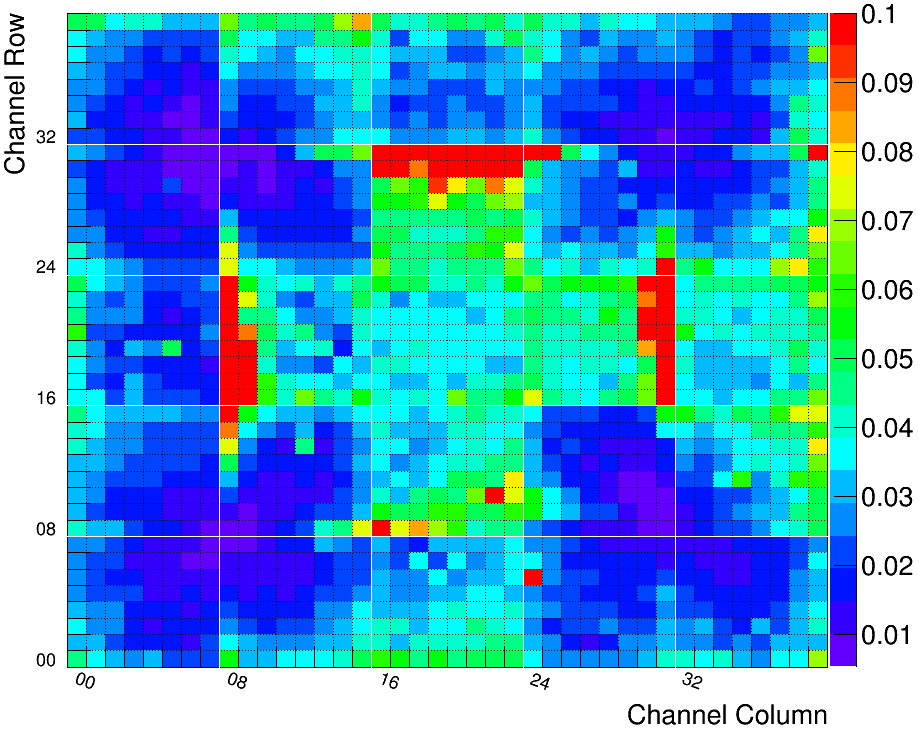}
\caption{The map of the relative error of gain for the high voltage setting of the normal data acquisition, where the maximum value of the Z-axis in the map is fixed at $10\%$.}\label{fig:error_of_gain}
\end{figure}

The measurement error on the gain of the 5 lower high voltage settings can be determined directly by the fitting error of Compton Edge. Whereas, for the high voltage setting of the normal scientific data acquisition the measurement error on the gain cannot be determined directly like the 5 lower high voltage settings, because the gain is indirectly calculated by Eq.~\eqref{equ:gain_vs_hv} with the measured parameters $p_0$ and $p_1$. However, the measurement error on the gain of the high voltage setting for the normal scientific data acquisition can be calculated using the standard error propagation equation and the results are shown in Figure~\ref{fig:error_of_gain}.

It can be clearly seen that the relative error of gain of those channels that are close to the $^{22}$Na sources is relatively small and of the order of $2\%$ due to large statistics. For those channels that are in the center and at the edge of the detector the relative error is a bit larger and of the order of $5\%$ due to a larger distance from the sources. It is also found that there are about 50 channels whose relative error is larger than $10\%$ as a result of low statistics, which is mainly because of the disadvantageous position of those channels to collect the coincident back to back Gamma-rays generated by the $^{22}$Na sources.

\newpage

\begin{figure}[!ht]
\centering
\subfigure[]{\includegraphics[height=5cm]{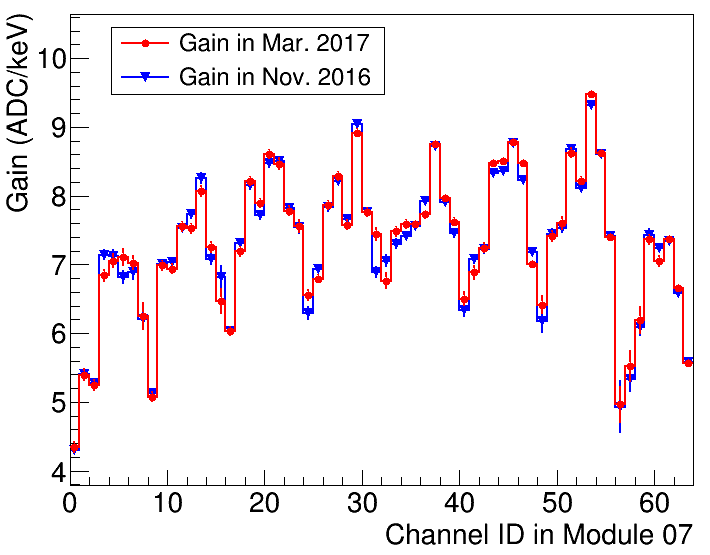}}
\hspace{2mm}
\subfigure[]{\includegraphics[height=5cm]{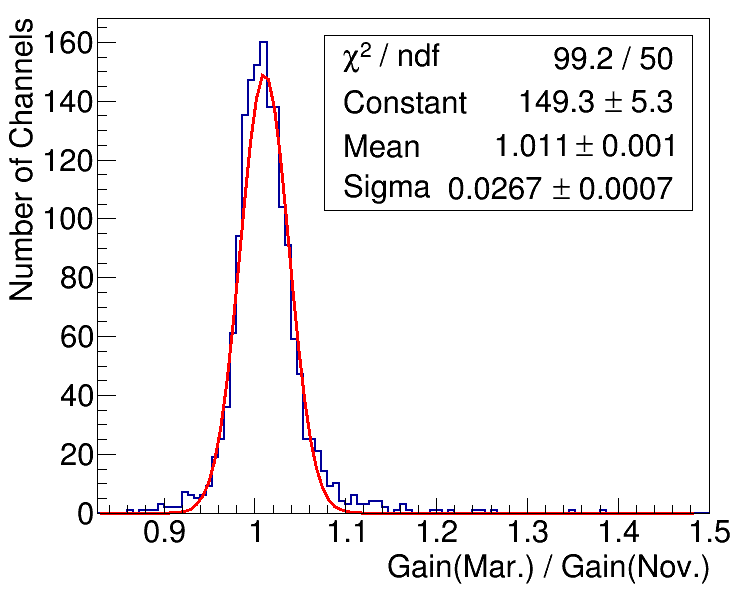}}
\caption{Comparison of the gain of a lower high voltage setting between November 2016 and March 2017. (a) shows the comparison for all the channels in module 07. (b) shows the distribution of the ratio of the gain in March 2017 for the lower high voltage setting over that in November 2016 for all 1600 channels.}\label{fig:gain_comp_Mar_Nov}
\end{figure}

Besides the data of the 5 lower high voltage settings acquired in November 2016, about 18 hours data of one of the 5 lower high voltage settings was taken in March 2017 for the purpose to check if there is any significant change of the gain during several months. Figure~\ref{fig:gain_comp_Mar_Nov} shows the result of the comparison for the gain between the two different times. Because there is a mean temperature difference of the module for a few degree Celsius between the two different times when the data was taken, the temperature correction was applied during the comparison for the gain using the mean slope of the gain change with temperature in the module level. Figure~\ref{fig:gain_comp_Mar_Nov}(a) shows the gain comparison for all the channels of one module, from which it can be seen that the gain of the two different times after temperature correction are very similar for most channels of the module. Figure~\ref{fig:gain_comp_Mar_Nov}(b) shows the distribution of the ratio of the gain in March 2017 for the lower high voltage setting over that in November 2016 for all 1600 channels, from which it can be seen that the mean ratio is very close to 1. However from the ratio distribution an average $1\%$ change was found which is possible to be a result of the not perfect temperature correction for gain. Even if this is a real gain change during the four months, it is in fact negligible, because the average relative measurement error on the gain for the lower high voltage setting is larger than $1\%$. Therefore, it is safe to use the gain parameters measured in November 2016 for the analysis of all the data acquired during the four months.

\section{Calibration induced error on polarization measurement}\label{sec:calib_sys_error}

As discussed in Section~\ref{sec:gain_hv}, the measured gain parameters of the normal data acquisition high voltage setting have non-negligible measurement errors for most channels due to limited statistics, even though all the in-orbit data of the 5 different high voltage settings has been used. In Section~\ref{sec:gain_temp}, it is also found that the gain parameters have an average $4\%$ relative change in the typical $5^\circ\mathrm{C}$ in-orbit temperature range of POLAR. Furthermore it is found that the temperature correction on gain is hard to perform for many channels resulting in an additional error. The gain is the most important calibration parameter to reconstruct the deposited energy of each bar. Therefore the systematic error on the polarization measurements for GRBs induced by the uncertainty from the gain dependence on high voltage and temperature needs to be estimated. This systematic error can be estimated using Monte Carlo simulations to study the variation of the $\mu_{100}$ of a typical GRB event induced by the error of gain parameters. For this purpose, the following steps of simulation were performed:

\begin{enumerate}[Step 1.]
\item Simulate a GRB event with $100\%$ polarization degree using a typical energy spectrum ($\alpha = -1.0, \beta = -2.5, E_{peak} = 200~\mathrm{keV}$) in the range 5-1000\,keV. The fluence is $\sim 1.6 \times 10^8\,m^{-2}$ with a plane circle area whose radius is $1\,m$ that can cover the full POLAR instrument and most of the parts of TG-2 around POLAR in the GEANT4 mass model. Three different groups of incident angles ($\theta, \phi$) are simulated, which are $(0^\circ, 0^\circ)$, $(30^\circ, 0^\circ)$, $(30^\circ, 45^\circ)$ respectively.
\item Perform 1000 times of digitization simulations using fixed gain parameters without any error for the purpose of acquiring the statistical error of $\mu_{100}$, which is defined as $\sigma_{stat}$.
\item Perform 1000 times of digitization simulations where the gain parameters are sampled within their error for each simulation, including the temperature induced error. The total error of $\mu_{100}$ is defined as $\sigma_{tot}$. Then the gain parameter induced systematic error on $\mu_{100}$ can be calculated by $\sigma_{sys}=\sqrt{\sigma_{tot}^2-\sigma_{stat}^2}$, and the relative systematic error is $\sigma_{sys} / \mu_{100}$, where $\mu_{100}$ is the mean value.
\end{enumerate}

Then the simulated data was analyzed using the same analysis pipeline as the real data to calculate $\mu_{100}$. The results of the simulation procedure for the three different groups of incident angles of the $100\%$ polarized GRB events are presented in Table~\ref{tab:gain_sys_error}. It can be seen that the gain parameters induced relative systematic error on $\mu_{100}$ for a typical GRB event is of the order of about $1.5\%$ according to the simulation. This error is significantly smaller than that found during the on-ground calibration of the instrument \cite{Kole2017}, where the systematic error induced by the poor energy calibration dominated the measurement uncertainty.

\begin{table}[!ht]
\centering
\caption{Gain parameters induced systematic error on $\mu_{100}$}\label{tab:gain_sys_error}
\begin{tabular}{|c|c|c|c|c|c|}\hline
$(\theta, \phi)$ & $\mu_{100}$ & $\sigma_{stat}$ & $\sigma_{tot}$ & $\sigma_{sys}$ & $\sigma_{sys} / \mu_{100}$ \\\hline
$(0^\circ, 0^\circ)$ & $0.360$ & $0.00175$ & $0.00521$ & $0.00491$ & $1.36\%$ \\\hline
$(30^\circ, 0^\circ)$ & $0.328$ & $0.00168$ & $0.00563$ & $0.00537$ & $1.64\%$ \\\hline
$(30^\circ, 45^\circ)$ & $0.325$ & $0.00165$ & $0.00492$ & $0.00464$ & $1.43\%$ \\\hline
\end{tabular}
\end{table}

Except for the gain parameter, the relative measurement error of other calibration parameters like pedestal level, ADC threshold, crosstalk factor, etc. can be easily reduced to a very small level ($<1\%$) using about 10 hours of in-orbit data. In order to estimate the level of systematic error on $\mu_{100}$ induced by other calibration parameters, the same simulation procedure was also performed for all other calibration parameters. It was found that the total relative systematic error on $\mu_{100}$ of a typical GRB event induced by other calibration parameters, when they are sampled within $1\%$ relative error in the simulation, is less than $0.2\%$. Therefore the total calibration induced systematic error on polarization measurement is dominated by the gain parameters. Furthermore, it can be concluded that the total relative systematic error on $\mu_{100}$ of a typical GRB event induced by the error of all calibration parameters is less than $2\%$. For future polarization analysis for a specific GRB event detected by POLAR, the same analysis of the calibration induced systematic error should be redone in detail according to the specific incident angle and energy spectrum of the GRB. It can however be assumed that the error will be of a similar level.


\section{Summary and Conclusions}\label{sec:5}

POLAR is a compact space-borne detector designed to perform reliable measurements of the polarization for transient sources like GRBs in the 50-500\,keV energy range. POLAR was successfully launched on-board the Chinese space laboratory TG-2 on 15th September in 2016 and successfully switched on afterwards. In order to reliably reconstruct the polarization information of the GRBs detected by POLAR from the in-orbit data sample and reduce the systematic effect of the instrument, a full detailed study of the in-orbit performance of the instrument was performed. All the related in-orbit calibration parameters of the instrument have been measured from the in-orbit data, including pedestal and noise level of the FEE, the nonlinearity function of the gain in the electronics, the ADC threshold of each channel, the crosstalk matrix of each module and the gain of each channel. The effect of temperature of the in-orbit environment of POLAR on all the calibration parameters was also carefully checked and studied, as well as the relationship between gain and high voltage for calculating the gain parameters corresponding to the normal data acquisition high voltage setting. All of those measured calibration parameters are sufficient and ready for POLAR to perform a reliable polarization analysis for the detected GRBs. Finally, in order to study the effect of the measurement error of calibration parameters on polarization measurements, the systematic error on $\mu_{100}$ of a typical GRB event induced by the measurement error of all the calibration parameters was studied by Monte Carlo simulation, and it was found that the total relative systematic error on $\mu_{100}$ of a typical GRB event induced by calibration parameters measurement error is less than $2\%$, which is expected to not dominate the measurement error on polarization degree for future polarization analysis for the detected GRBs of POLAR.

\section{Acknowledgments}

We gratefully acknowledge the financial support from the National Basic Research Program (973 Program) of China (Grant No. 2014CB845800), the Joint Research Fund in Astronomy under the cooperative agreement between the National Natural Science Foundation of China and the Chinese Academy of Sciences (Grant No. U1631242), the National Natural Science Foundation of China (Grant No. 11503028, 11403028), the Strategic Priority Research Program of the Chinese Academy of Sciences (Grant No. XDB23040400), the Swiss National Science Foundation, the Swiss Space Office (ESA PRODEX program) and the National Science Center of Poland (Grant No. 2015/17/N/ST9/03556).

\newpage

\bibliographystyle{unsrt}
\bibliography{references}

\end{document}